\documentclass[preprint,nofootinbib,aps,superscriptaddress,eqsecnum]{revtex4-1}
\usepackage{float}
\usepackage{graphicx}
\usepackage{amsmath}
\usepackage{amsfonts}
\usepackage{amssymb}
\usepackage[bookmarks=true,bookmarksopen=true,
bookmarksnumbered=true,
colorlinks=true,linkcolor=black,
citecolor=red]{hyperref}
\usepackage{subfigure}
\usepackage{slashed}
\usepackage{setspace} 
\usepackage{ulem}
\usepackage{caption}
\usepackage{color}
\allowdisplaybreaks
\def\bea{\begin{eqnarray}}
	\def\eea{\end{eqnarray}}
\def\be{\begin{equation}}
	\def\ee{\end{equation}}
\def\nn{\nonumber}

\begin{document}
	
	%\vspace{-0.2cm}
	\title{{Flavoring the production of Higgs pairs}}
	
	\author{\small M. A. Arroyo-Ure\~na }
	\email{marco.arroyo@fcfm.buap.mx}
	\affiliation{\small Facultad de Ciencias F\'isico-Matem\'aticas, Benem\'erita Universidad Aut\'onoma de Puebla, C.P. 72570, Puebla, M\'exico,}
	\affiliation{\small Centro Interdisciplinario de Investigaci\'on y Ense\~nanza de la Ciencia (CIIEC), Benem\'erita Universidad Aut\'onoma de Puebla, C.P. 72570, Puebla, M\'exico.}

	\author{\small J. Lorenzo D\'iaz-Cruz }
	\email{jldiaz@fcfm.buap.mx }
	\affiliation{\small Facultad de Ciencias F\'isico-Matem\'aticas, Benem\'erita Universidad Aut\'onoma de Puebla, C.P. 72570, Puebla, M\'exico,}
	\affiliation{\small Centro Interdisciplinario de Investigaci\'on y Ense\~nanza de la Ciencia (CIIEC), Benem\'erita Universidad Aut\'onoma de Puebla, C.P. 72570, Puebla, M\'exico.}

	\author{E. A. Herrera-Chac\'on}
	\email{edwin.a.herrera-chacon@durham.ac.uk}
	\affiliation{Institute for Particle Physics Phenomenology, Durham University, South Road, Durham, DH1 3LE}

	\author{\small T. A. Valencia-P\'erez }
	\email{tvalencia@fisica.unam.mx}
	\affiliation{\small{Instituto de F\'isica, 
			Universidad Nacional Aut\'onoma de M\'exico, C.P. 01000, CDMX, M\'exico.}}

	\author{\small J. Mejia Guisao.}
	\email{jhovanny.andres.mejia.guisao@cern.ch}
	\affiliation{\small Departamento de Física, Centro de Investigación y de Estudios Avanzados del IPN,}
	\affiliation{\small Centro Interdisciplinario de Investigaci\'on y Ense\~nanza de la Ciencia (CIIEC), Benem\'erita Universidad Aut\'onoma de Puebla, C.P. 72570, Puebla, M\'exico.}
	
	\hspace*{-2cm}
	\begin{abstract}
		{\small We present a study on the possibility of observing a hypothetical particle known as the Flavon $H_F$, which is predicted in an extension of the standard model that includes the so-called Froggatt-Nielsen mechanism. The proposed decay channel is through a $b\bar{b}h$ final state, where the Higgs boson $(h)$ decays to a pair of photons or a pair of $b$ quarks $(h\to \gamma\gamma,\,b\bar{b})$. We found that, under special scenarios of the model parameter space, the processes analyzed could provide evidence for the existence of the Flavon in the next stage of the LHC: the High Luminosity LHC. Specifically, we predict a \textit{signal significance} of $5\sigma$ ($2\sigma$) in the $h\to b\bar{b}$ ($h\to\gamma\gamma$) channel for a Flavon mass of $800$ ($900$) GeV and an integrated luminosity of $2500$ ($3000$) fb$^{-1}$.
		}
	\end{abstract}

	\maketitle

	\section{Introduction}
\begin{figure}[!hbt]
		\includegraphics[width=10cm]{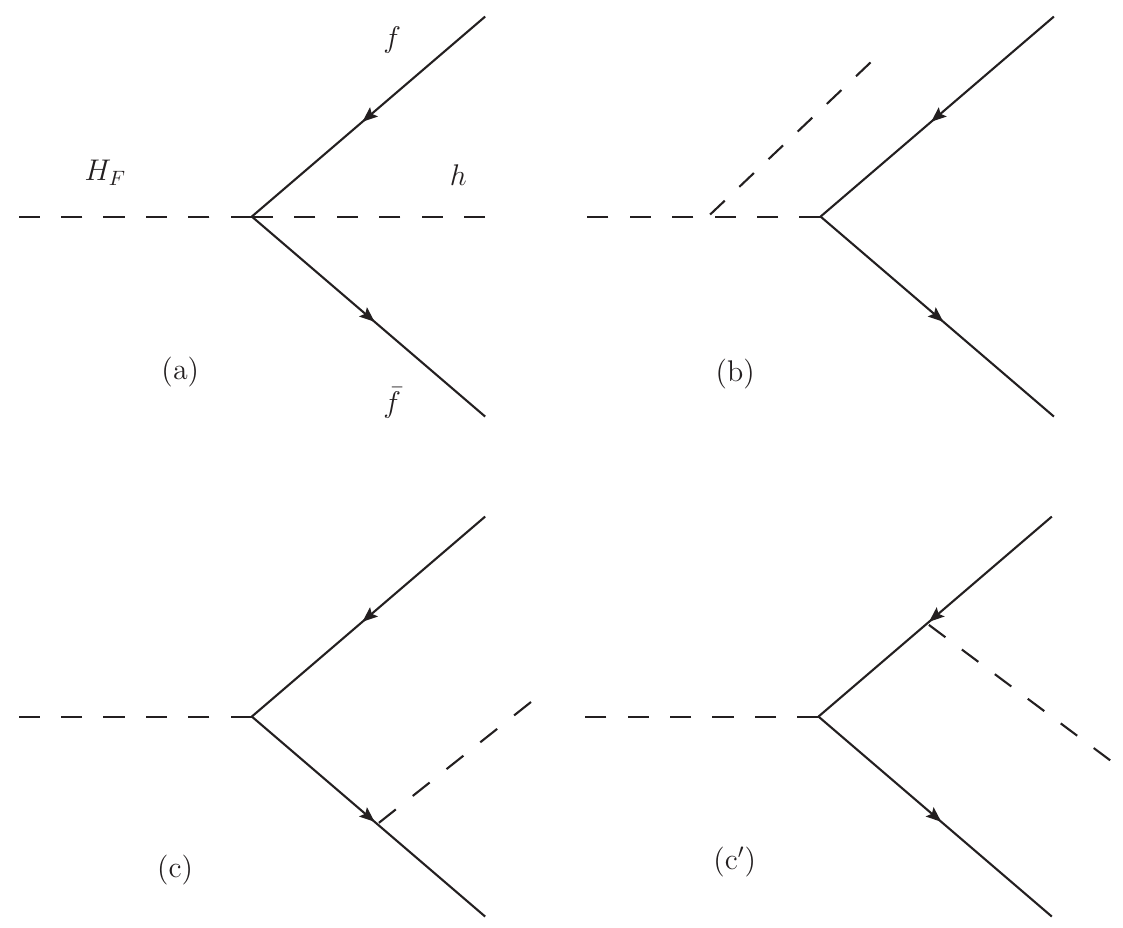}
		\caption{Feynman diagrams inducing the $H_F\to \bar{f}fh$ decay in the FNSM.\label{Htoffh}}
	\end{figure}

	After the discovery of the Higgs boson by the ATLAS and CMS collaborations \cite{ATLAS:2012yve, CMS:2012qbp} at the Large Hadron Collider (LHC), the particle physics community has focused on understanding the interactions of the Higgs boson with other particles, including itself. In the last years, the ATLAS and CMS collaborations have reported results on searches for di-Higgs proton-proton ($pp$) production in the channels $pp\to hh,\, (h\to b\bar{b},\,h\to b\bar{b}),\,(h\to b\bar{b},\,h\to\gamma\gamma),\,(h\to b\bar{b},\,h\to WW^*),\,(h\to b\bar{b},\,h\to ZZ^*),\,(h\to b\bar{b},\,h\to \tau^-\tau^+)$ \cite{ATLAS:2023qzf, ATLAS:2024lsk, ATLAS:2023gzn,ATLAS:2024yuv, ATLAS:2023elc}. These searches have studied the different ways in which the Higgs boson is produced, such as ggh, VBF, in association with top-quark pairs, etc.  We notice that in all channels the $b\bar{b}$ pair prevails, motivated by these final states we undertake an analysis of production and decay of a scalar (so-called Flavon, denoted by $H_F$) predicted in a model that implements the Froggatt-Nielsen (FN) mechanism \cite{Froggatt:1978nt}, which we will refer to as the FN Singlet Model (FNSM). Such a model assumes that above some scale $\Lambda$, there is a symmetry (perhaps of Abelian type $U(1)_F$) that forbids the appearance of Yukawa couplings; SM fermions are charged under this symmetry. However, Yukawa matrices can arise through non-renormalizable operators. The Higgs spectrum of these models includes the flavon, which could mix with the Higgs bosons when the flavor scale is of the order of the TeVs. The flavon phenomenology has been studied by several authors, including the production and decay processes $pp\to H_F\to hh (h\to \gamma\gamma, \,h\to b\bar{b})$, $pp\to H_F\to ZZ (Z\to \ell^-\ell^+)$, $pp\to H_F\to tc (t\to \ell\nu_{\ell}b)$, $pp\to H_F\to \tau\mu (\tau\to \ell\nu_{\tau}\nu_{\ell})$ ($\ell= e,\,\mu$)~\cite{Bauer:2016rxs, Barradas-Guevara:2017ewn, Arroyo-Urena:2022oft, Arroyo-Urena:2018mvl, Arroyo-Urena:2019fyd}. In this study, we explore a final state in which current experimental reports on searches for the di-Higgs boson can shed light on the signal we propose, i.e., the production of the flavon via proton-proton collisions and later it decays into $b\bar{b}h$ for two particular channels, namely $h\to b\bar{b}$ and $h\to\gamma\gamma$. The induced Feynman diagrams of the decay $H_F\to f\bar{f}h$ in the theoretical framework of the FNSM are presented in Fig. \ref{Htoffh}\footnote{The diagram $(b)$ has a scalar propagator in which both the Higgs boson $h$ and the flavon $H_F$ propagate.}.
	The main difference between the di-Higgs boson production signatures and our signal is the resonant effect that offers the decay $H_F\to b\bar{b}h$. This is a great advantage from an experimental point of view.

	The paper is organized as follows. Section ~\ref{se:model} presents the relevant aspects of the model, including expressions for the masses and the Feynman rules to be used in the subsequent analysis.
	Afterwards, both theoretical and experimental constraints that have a direct impact on the predictions are studied in Sec.~\ref{se:conts}. Meanwhile, Sec. \ref{se:col_an} is focused on the analysis of the signal, $pp\to H_F\to hb \bar{b} $ $(h\to b \bar{b},\,\gamma\gamma)$ as well as its background processes. A multivariate analysis is also included. Finally, we present our conclusions in Sec.~\ref{se:concl}.

	\section{The model} 
	\label{se:model}
	
	\subsection{The scalar sector} 
	The FNSM's scalar sector add one complex singlet FN scalar $S_F$ to the SM, namely,
	%%%%%%%%%%%
	%%%%%%%%%%%
	%%%%%%%%%%%
	\begin{eqnarray} 
		%	& \Phi = \left( \begin{array}{  c} 0 \\ \frac{  v + \phi^0}{\sqrt 2}\\
			%	\end{array}  \right), \label{dec_doublets}&\\ 
		& S_F = \frac {(v_s + S_R + i S_I )}{ \sqrt 2 }   , \label{dec_Singlet} &
	\end{eqnarray}
	%%%%%%%%%%%%%%%%%%%
	%%%%%%%%%%%
	%%%%%%%%%%%
	where $v_s$ represents the vacuum expectation value (VEV) of the FN singlet. The scalar potential is invariant under the FN $U(1)_F$ flavor symmetry. Under this flavor symmetry, $S_F$ and the SM Higgs doublet $\Phi=(\frac{  v + \phi^0}{\sqrt 2}\,\,\, \phi^+)^T$ transform as $S_F \to e^{i\alpha} S_F$ and $\Phi \to \Phi$, respectively. We note that such a scalar potential allows a complex VEV, $\langle S_F\rangle_0=\frac{v_s}{\sqrt{2}}e^ {i\alpha}$, but in this work we assume the scenario in which the Higgs potential is CP-conserving ($\alpha = 0$).
	Then, the CP-conserving Higgs potential reads:
	%%%%%%%%%%%
	%%%%%%%%%%%
	%%%%%%%%%%%
	\begin{eqnarray} \label{potential} 
		V_0=-\frac {1}{2} m_1^2\Phi^ \dagger \Phi-\frac{1}{2} m_{2}^2 
		S_F^*S_F +\frac {1}{2} \lambda_1 \left(\Phi^ \dagger \Phi\right)^2+\lambda_2
		\left(S_F^*S_F\right)^2
		+\lambda_ {3} \left(\Phi^ \dagger \Phi\right)\left(S_F^* S_F\right). 
	\end{eqnarray}
	%%%%%%%%%%%
	%%%%%%%%%%%
	%%%%%%%%%%%
	After the spontaneous symmetry breaking of the $U(1)_F$ flavor symmetry by the VEVs of the spin-0 fields $(\Phi, S_F)$, a massless Goldstone boson will emerge in the physical spectrum. To give a mass to it, we embed a soft $U(1)_F$ breaking term in the scalar potential: 
	%%%%%%%%%%%
	%%%%%%%%%%%
	%%%%%%%%%%%
	\begin{eqnarray}
		V_{\rm soft} = -\frac{m_3^2}{2} \left (S_F^{2} + S_F^{*2} \right).  
	\end{eqnarray}
	Thus, the full scalar potential is given by:
	\begin{eqnarray}
		V = V_0 + V_{\rm soft}. 
	\end{eqnarray}
	%%%%%%%%%%%
	%%%%%%%%%%%
	%%%%%%%%%%%
	The parameter $\lambda_3$ in Eq.\eqref{potential} allows a mixing between the
	Flavon and the Higgs fields after both the $U(1)_F$ flavor and EW symmetries
	are spontaneously broken, generating the masses of the Flavon and Higgs field, as
	presented below while the pseudoscalar Flavon $(S_I)$ mass is generated via the soft $U(1)_F$ flavor symmetry breaking term $V_{\rm soft}$.
	Once the minimization of the potential $V$ is done, we obtain relations between the parameters of $V$ as follows:
	\begin{eqnarray} 
		m_{1} ^2  &=&  v^2 \lambda_1 + v_s^2 \lambda_{3},   \\
		m_{2} ^2 &=& -2 m^2_{3} + 2 v_s^2 \lambda_2 + v^2 \lambda_{3}.
	\end{eqnarray}
	%%%%%%%%%%%
	%%%%%%%%%%%
	%%%%%%%%%%% 
	Since all parameters of the scalar potential are assumed to be real, the imaginary and real parts of $V$ do not mix. The CP-even mass matrix can be written in the $(\phi_0, S_R)$ basis as:
	%%%%%%%%%%%
	%%%%%%%%%%%
	%%%%%%%%%%% 
	\begin{equation} 
		M^2_S =
		\left( \begin{array}{cc} 
			\lambda_1 v^2      &  \lambda_{3} v v_s \\
			\lambda_{3}v v_s   &  2 \lambda_2 v_s^2
		\end{array}  \right),
	\end{equation} 
	%%%%%%%%%%%
	%%%%%%%%%%%
	%%%%%%%%%%%
	whose mass eigenstates can be obtained through the following $2\times 2$ rotation:
	%%%%%%%%%%%
	%%%%%%%%%%%
	%%%%%%%%%%%
	\begin{eqnarray} 
		\phi^0   &=&\cos\alpha \  h + \sin\alpha  \  H_F,   \\
		S_R    &=& -\sin\alpha \  h + \cos\alpha \  H_F, 
	\end{eqnarray}
	%%%%%%%%%%%
	%%%%%%%%%%%
	%%%%%%%%%%% 
	where $\alpha$ is the mixing angle. We identify $h$ with the SM-like Higgs boson with mass $M_h$ = 125.5 GeV. Meanwhile, the mass eigenstate $H_F$ is identified 
	with the CP-even Flavon. The CP-odd Flavon $A_F \equiv S_I$ will obtain its mass from the $V_{\rm {soft}}$ term such that $M^2_{A_F} = 2m_3^2 $. 
	We will work (as free parameters) with the physical masses $M_{S}\, (S=h,\,H_F,\,A_F)$ and the mixing angle $\alpha$, whose relations with the quartic couplings of the scalar potential in Eq.~(\ref{potential}) read:
	%%%%%%%%%%%
	%%%%%%%%%%%
	%%%%%%%%%%%
	%%%%%%%%%%%%
	\begin{eqnarray}\label{eq:relate}
		\lambda_1&=& \frac{ \cos\alpha^2 M_h^2+\sin\alpha^2 M_{{H_F}}^2}{v^2},\nn\\
		\lambda_2&=& \frac{M_{{A_F}}^2+{\cos\alpha }^2 M_{{H_F}}^2+{\sin\alpha }^2 M_h^2}{2 v_s^2},\\
		\lambda_3&=& \frac{ \cos\alpha \, \sin\alpha }{ v v_s} \, ( M_{{H_F}}^2 -  M_h^2).\nn
	\end{eqnarray} 
	%%%%%%%%%%%%%%
	%%%%%%%%%%%
	%%%%%%%%%%%
	%%%%%%%%%%%
	
	%%%%%%%%%%%%%%%%%%%
	\subsection{The Yukawa sector} 
	%%%%%%%%%%%%%%%%%%%%%%
	The effective $U(1)_{F}$ invariant Yukawa Lagrangian includes terms that become the Yukawa couplings after the $U(1)_{F}$ flavor symmetry is spontaneously broken. It is given by \cite{ Froggatt:1978nt}:
	%%%%%%%%%%%
	%%%%%%%%%%%
	%%%%%%%%%%%
	\begin{align} 
		\mathcal{L}_ Y &= \rho^d_{ ij } \left( \frac{S_F}{\Lambda} 
		\right)^{q_{ ij }^d}  \bar{Q}_i d_j  \tilde \Phi 
		+ \rho^u_{ ij } \left(\frac{S_F}{\Lambda}\right)^{q_{ij }^u}\bar{Q}_i u_j 
		\Phi \nonumber\\&+ \rho^\ell_{ij}\left(\frac{S_F}{\Lambda}\right)^{q_{ij}^l}
		\bar{L}_i \ell_j \Phi  + \rm h.c.,
		\label{eq:fermlag}  
	\end{align} 
	%%%%%%%%%%%
	%%%%%%%%%%%
	%%%%%%%%%%%
	where $\rho^{f}_{ij}$ ($f=u,\,d,\,\ell$) are dimensionless parameters ostensibly of order $\mathcal{O}(1)$. The amounts $q_{ij}^f$ represent the Abelian charges that reproduce the observed fermion masses and $\Lambda$ is the ultraviolet mass scale, which is not predicted by the Froggat-Nielssen mechanism, it can be between the weak and the Planck scale. However, there is an essential requirement: the flavor symmetry must be broken in a way such that the ratio $\frac{v_s}{\sqrt{2}\Lambda}\lesssim 1$. To generate the Yukawa couplings from Lagrangian \eqref{eq:fermlag} one must spontaneously break both, the $U(1)_{F}$ as well as EW symmetries. Once this is done, we arrive to the $S f_i\bar{f}_i$ interactions as shown in Table \ref{Diagonal_couplings}. We note that to avoid large deviations from the SM coupling, $v_s\approx\Lambda$ and $\cos\alpha\approx -1$ are required, as below.
	\begin{table}
		
		\caption{Diagonal $S XX$ interactions, ($S=h,\,H_F,\,A_F$).}\label{Diagonal_couplings}
		
		\begin{centering}
			\begin{tabular}{c c}
				\hline 
				Vertex $SXX$ & Coupling\tabularnewline
				\hline 
				\hline 
				$hf_i\bar{f}_i$ & $\frac{v_{s}m_{f}}{v\Lambda^{2}}\Big(v\sin\alpha-v_{s}\cos\alpha\Big)$\tabularnewline
				\hline 
				$hZZ$ & $g\frac{m_{Z}}{c_{W}}\cos\alpha$\tabularnewline
				\hline 
				$hWW$ & $gm_{W}\cos\alpha$\tabularnewline
				\hline 
				$H_{F}f_i\bar{f}_i$ & $-\frac{v_{s}m_{f}}{v\Lambda^{2}}\Big(v_{s}\sin\alpha+v\cos\alpha\Big)$\tabularnewline
				\hline 
				$H_{F}ZZ$ & $g\frac{m_{Z}}{c_{W}}\sin\alpha$\tabularnewline
				\hline 
				$H_{F}WW$ & $gm_{W}\sin\alpha$\tabularnewline
				\hline 
				$A_{F}f_i\bar{f}_i$ & $-\frac{v_{s}m_{f}}{\Lambda^{2}}$\tabularnewline
				\hline 
				$A_{F}ZZ$ & $0$\tabularnewline
				\hline 
				$A_{F}WW$ & $0$\tabularnewline
				\hline 
			\end{tabular}
			\par\end{centering}
	\end{table}

	The CP-even $H_F$ and the CP-odd $A_F$ flavons are assumed to be heavier than $h$.
	
	Meanwhile, to induce non-diagonal $S f_i\bar{f}_j$ interactions, we proceed as described below. By considering the unitary gauge one can make the following first order expansion of the neutral component of the heavy Flavon field $S_F$ around its VEV $v_s$:
	%%%%%%%%%%%
	%%%%%%%%%%%
	%%%%%%%%%%%
	\begin{align} 
		\Bigg(\frac{S_F}{\Lambda}\Bigg)^{q_{ ij }} &=\left(\frac{v_s+S_R+iS_I}{  \sqrt 2\Lambda}  \right)^{q_{ ij }} \nonumber\\&
		\simeq \left(\frac{v_s}{ \sqrt 2\Lambda}  \right)^{q_{ ij }} \left[1+q_{ ij }\left(\frac{S_R+iS_I}{v_s}\right)\right],
	\end{align}
	%%%%%%%%%%%
	%%%%%%%%%%%
	%%%%%%%%%%% 
	which leads to the following Lagrangian interaction after replacing the mass eigenstates:
	%%%%%%%%%%%
	%%%%%%%%%%%
	%%%%%%%%%%%
	\begin{eqnarray} \label{Yukalagrangian} 
		\mathcal {L}  _Y &=& \frac 1  v [\bar{U}  M^u U+\bar{D}  M^d D+\bar{L} M^ \ell L](c_ \alpha h+s_ \alpha H_F) \nonumber\\
		&+&\frac{v }{ \sqrt 2 v_s } [\bar{U}_i\tilde Z_{ij}^u U_j+\bar{D}_i\tilde Z_{ij}^d D_j+\bar{L}_i\tilde Z_{ij}^ \ell  L_j]\nonumber\\&\times&
		(-\sin\alpha h+\cos\alpha H_F+iA_F)+ \rm h.c.,  
	\end{eqnarray} 
	%%%%%%%%%%%
	%%%%%%%%%%%
	%%%%%%%%%%%
	Here, $M^f$ denotes the diagonal fermion mass matrix. We encapsulate the Higgs-Flavon couplings in the $\tilde{Z}_{ij}^f=U_L^f Z_{ij}^f U_L^{f\dagger}$ matrices. In the flavor basis, the $Z_{ij}^f$ matrix elements are given by:
	%%%%%%%%%%%
	%%%%%%%%%%%
	%%%%%%%%%%%
	\begin{equation}
		Z_{ij}^f= \rho_{ij}^f \left(\frac{v_s}{\sqrt 2\Lambda}
		\right)^{q_{ij}^f}q_{ij}^f,
	\end{equation}
	%%%%%%%%%%%
	%%%%%%%%%%%
	%%%%%%%%%%%
	which (in general) remains non-diagonal even after diagonalizing the mass matrices, giving rise to FV interactions. 
	
	Finally, in addition to the Yukawa couplings, we also need the $H_Fhf_i\bar{f}_j$ couplings for our calculations. In the FNSM these interactions are given by
	\begin{equation}
		H_Fhf\bar{f}=\frac{m_f v_s}{\sqrt{2}\Lambda^2}(1-2\cos^2\alpha).
	\end{equation} 
	As a particular case we will explore the scenario where $f=b$. This choice is motivated by experimental reports on Higgs pair searches \cite{ATLAS:2023qzf, ATLAS:2024lsk, ATLAS:2023gzn,ATLAS:2024yuv, ATLAS:2023elc}
	%%%%%%%%%%%
	%%%%%%%%%%%
	%%%%%%%%%%%
	%%%%%%%%%%%
	%%%%%%%%%%%
	
	%%%%%%%%%%%%%%%%%%%%%%%%%%%%%%%%%%%%%%%%%%%%
	
	%%%%%%%%%%%%%%%%%%%%%%%%%%%%%%%%%%%%%%%%%%%%%%%%%%%
	
	%%%%%%%%%%%%%%%%%%%
	\section{Constraints on the FNSM parameter space}
	\label{se:conts}
	%%%%%%%%%%%%%%%%%%%%%%%

	In order to compute a realistic numerical analysis of the signals proposed in this project, i.e., $pp\to H_F\to hb\bar{b}~(h\to b \bar{b},\, \tau^-\tau^+,\, WW^*,\,\gamma\gamma,\,ZZ^*)$, we need to constrain the free FNSM parameters involved in the upcoming calculations, namely,
	\begin{itemize}
		\item The mixing angle $\alpha$ of the real components of the doublet $\Phi$ and the FN singlet $S$,
		\item FN singlet VEV $v_s$,
		\item The ultraviolet mass scale $\Lambda$,
		\item CP-even scalar mass $M_{H_F}$.
	\end{itemize} 
	
	These parameters can be constrained by several types of theoretical restrictions like absolute vacuum stability, triviality, perturbativity, and unitarity of scattering matrices and different experimental data, mainly, LHC Higgs boson data upper limits on the production cross-section of additional Higgs states. We also consider bounds on Lepton Flavor Violating processes (LFVp) $L_i\to \ell_j\ell_k\bar{\ell}_k$, $\ell_i\to \ell_j\gamma$. Measurements on $\mathcal{BR}(B_{s,\,d}^0\to\mu^+\mu^-)$ and the anomalous magnetic moments of the muon and electron $\Delta a_{\mu}$ and $\Delta a_{e}$, respectively, are also presented. Finally, we also take into account quark flavor constraints: $B-\bar{B}$, $K-\bar{K}$ and $D-\bar{D}$ mixing.
	
	\subsection{theoretical constraints}
	\subsubsection{Stability of the scalar potential}
	It is necessary to control the stability of the scalar potential in Eq. \eqref{potential}, it should be bounded from below, i.e., it should not approach negative infinity along any direction of the field space ($h,H_F,A_F$) at large field values. The quadratic terms in the scalar potential are very suppressed compared to the quartic terms. So, the absolute stability conditions read \cite{Khan:2014kba}:
	%%%%%%%%%%%
	%%%%%%%%%%%
	%%%%%%%%%%%
	\begin{equation}
		\Big(\lambda_1 ,\, \lambda_2 ,  \, \lambda_3 + \sqrt{2 \lambda_1 \lambda_2} \Big)> 0.
	\end{equation}
	%%%%%%%%%%%
	%%%%%%%%%%%
	%%%%%%%%%%%
	The quartic couplings are evaluated at a scale $\Lambda$ using the Renormalization Group Evolution (RGE) equations. If the scalar potential in Eq.~(\ref{potential}) has a metastable EW vacuum, then these conditions should be modified \cite{Khan:2014kba}. To constrain the scalar field masses $M_{\phi}$, the VEV of the complex singlet $v_s$ and the mixing angle $\alpha$, we can use Eqs. \eqref{eq:relate} to translate these limits into those on the model parameters.

	\subsubsection{Perturbativity and unitarity constraints}\label{theoretical_constraints}
	
	%%%%%%%%%%%%%%%%%%%%%%%%%%%%%%%%%%%%%%%%%%%%
	%\begin{figure}[!t]
	%	\begin{center}
		%		\includegraphics[scale=0.155]{LamSPlot.png}
		%		\includegraphics[scale=0.155]{Lam11Plot.png}
		%		\includegraphics[scale=0.155]{LamUPlot.png}
		%	\end{center}
	%	\caption{ In the first two plots we show the perturbative bounds on the quartic couplings $\lambda_{2,3}$ while the third plot shows the stringent unitary bounds on $\lambda_U$.}
	%	\label{PLimit}
	%\end{figure}
	%%%%%%%%%%%%%%%%%%%%%%%%%%%%%%%%%%%%%%%%%%%%%%%%%%%
	
	%%%%%%%%%%%%%%%
	The upper limits are presented in Eq. \eqref{UL4pi} are necessary to ensure that the radiatively corrected scalar potential of the FNSM remains perturbative at any energy scale:
	%%%%%%
	%%%%%%%%%%%
	%%%%%%%%%%%
	%%%%%%%%%%%
	\begin{equation}\label{UL4pi}
		\mid \lambda_1, \lambda_2, \lambda_3\mid \leq 4 \pi.
	\end{equation}
	%%%%%%%
	%%%%%%%%%%%
	%%%%%%%%%%%
	%%%%%%%%%%%
	%%%%%%
	
	The quartic couplings that come from the scalar potential are also highly constrained by the unitarity of the $S$-matrix. Even at large field values, one can obtain the $S$-matrix by considering several $(P)S-(P)S$, $V-V$ and $(P)S-V$ interactions in $2\to2$ body processes, where $P(S,\,V)$ stands for a pseudo-scalar boson (scalar, gauge boson). The unitarity of the $S$-matrix demands that its eigenvalues be less than $8\pi$ \cite{Cynolter:2004cq,Khan:2014kba}. Within the theoretical framework of FNSM, using the equivalence theorem, the unitary bounds are given by:
	%%%%%%%%%%%
	%%%%%%%%%%%
	%%%%%%%%%%% 
	%%%%%%
	\begin{eqnarray}\label{UnitaryBounds}
		\lambda_1 \leq 16 \pi  \quad {\rm and} \quad \Big| {\lambda_1}+{\lambda_2} \pm \sqrt{ ({\lambda}_1-{\lambda_2})^2+(2/3 \lambda_3)^2}\Big|\equiv |\lambda_U^{\pm}| \leq 16/3 \pi.
	\end{eqnarray}
	%%%%%%%%%%%
	%%%%%%%%%%%
	%%%%%%%%%%%
	%%%%%%%%%%%%% 
	By using the Eqs. \eqref{eq:relate}, \eqref{UL4pi} and \eqref{UnitaryBounds}, we can constrain the scalar singlet VEV $v_s$, the heavy Higgs masses ($M_{H_F}$, $M_{A_F}$) and the mixing angle $\alpha$. 
	
\begin{figure}[htb!]
		\begin{center}
			\includegraphics[scale=0.25]{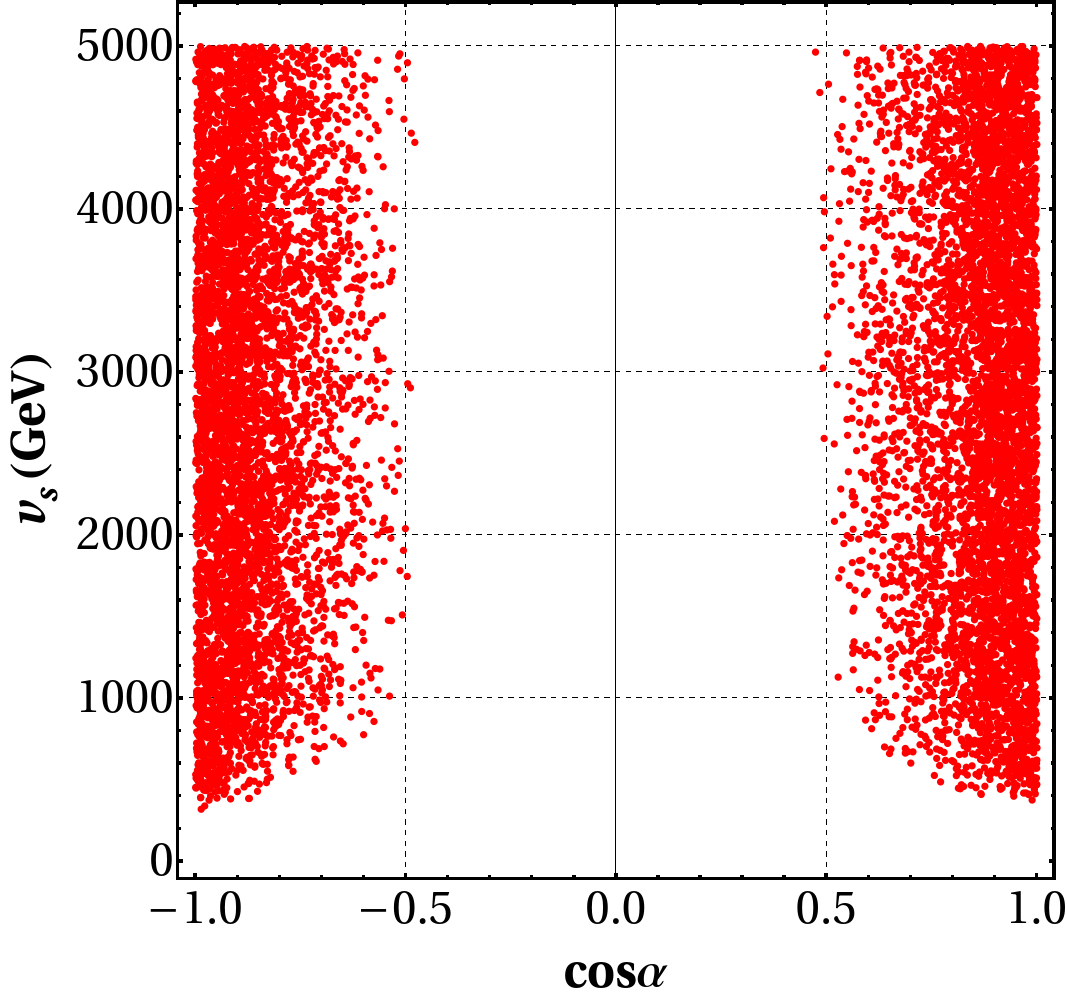}
		\end{center}
		\caption{VEV of the FN singlet $v_s$ as a function of cosine of the mixing angle $\alpha$. The red points indicate those allowed by all theoretical restrictions as mentioned in the main text.}
		\label{PLimit}
	\end{figure}

	Figure \ref{PLimit} displays the $\cos\alpha-v_s$ plane whose points represent those allowed by the theoretical constraints, perturbativity and unitarity of the $S$-matrix. For this purpose we generate random points such that they meet the relations \eqref{eq:relate}, \eqref{UL4pi} and \eqref{UnitaryBounds}. We present in Table \ref{RPTC} the scanned intervals to achieve that proposal.
	
	\begin{table}
		
		\caption{Scanned parameters.}\label{RPTC}
		
		\begin{centering}
			\begin{tabular}{|c|c|}
				\hline 
				Parameter & Interval\tabularnewline
				\hline 
				\hline 
				$\cos\alpha$ & {[}-1, 1{]}\tabularnewline
				\hline 
				$M_{H_{F}}$ & 800-1500 GeV\tabularnewline
				\hline 
				$M_{A_{F}}$ & 800-1500 GeV\tabularnewline
				\hline 
				$v_{s}$ & $v$-5000 GeV\tabularnewline
				\hline 
			\end{tabular}
			\par\end{centering}
	\end{table}
	
	We find that $|\lambda_U^+| \leq 16\pi/3 $ is the most stringent upper bound for the scalar quartic couplings, which is transferred to a lower limit on the scalar singlet VEV $v_s\geq (276,\,345,\,415,\,519)$ GeV, for specific masses of $M_{H_F}=M_{A_F}= (800,\,1000,\,1200,\,1500)$ GeV and $\cos\alpha=-0.995$. Note that we are working at the EW scale only, as detailed RGE analysis is beyond the scope of this work. We also observe a greater density around $1$ and $-1$, this is because $\alpha$ must tend to zero so as not to have large deviations from the $hf\bar{f}$ coupling of the SM.
	
	It is important to highlight that the masses were analyzed in the range shown in Table \ref{RPTC} because in a previous study by one of us (and others)\cite{Arroyo-Urena:2022oft}, it was found that $M_{H_F}>800$ GeV to release the limit on the cross-section of the process $pp\to S\to hh$ reported by the ATLAS Collaboration \citep{ATLAS:2021ifb}, where $S$ is a resonant (pseudo-)scalar particle. 
	
	\subsection{Experimental constraints}
	\subsubsection{LHC Higgs boson data}
	
	We complement the theoretical constraints by using experimental measurements of Higgs boson physics \cite{ParticleDataGroup:2024cfk}, specifically we consider the \textit{signal strengths} defined as
	\begin{equation}
		\mathcal{R}_{X}=\frac{\sigma(pp\to h)\cdot\mathcal{BR}(h\to X)}{\sigma(pp\to h^{\text{SM}})\cdot\mathcal{BR}(h^{\text{SM}}\to X)},
	\end{equation}
	where $\sigma(pp\to H_i)$ is the production cross-section of $H_i$, with $H_i=h,\,h^{\text{SM}}$; here $h$ is the SM-like Higgs boson coming from an extension of the SM and $h^{\text{SM}}$ is the SM Higgs boson; $\mathcal{BR}(H_i\to X)$ is the branching ratio of the decay $H_i\to X$, with $X=b\bar{b},\,c\bar{c},\;\tau^-\tau^+,\;\mu^-\mu^+,\;WW^*,\;ZZ^*,\;\gamma\gamma$.
	\subsubsection{Lepton Flavor Violating processes}
\begin{figure}[htb!]
		\begin{center}
			\includegraphics[scale=0.4]{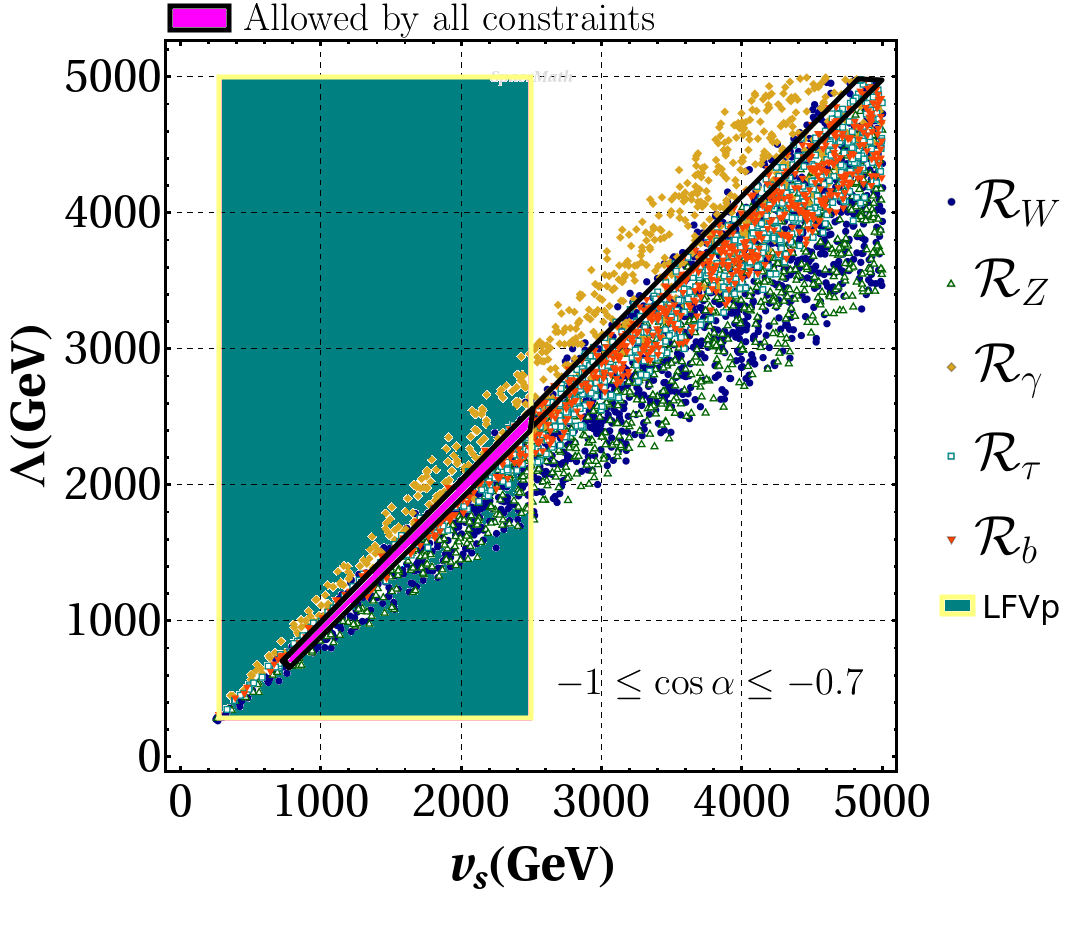}
		\end{center}
		\caption{Ultraviolet mass scale $\Lambda$ as a function of the VEV of the complex singlet $v_s$. The different points represent individual $\mathcal{R}_X's$, while the enclosed area by solid black lines is the common zone of them. The green rectangle stands for the allowed region by LFVp. Meanwhile, the cyan area is the result of applying all discussed theoretical and experimental constraints (intersection).}
		\label{fig:RXs}
	\end{figure}

	Furthermore, we also analyze several LFVp that can help us even more in constraining the parameters involved in subsequent calculations. Specifically, we use i) upper limits on $\mathcal{BR}(L_i\to \ell_j\ell_k\bar{\ell}_k)$ and $\mathcal{BR}(\ell_i\to \ell_j\gamma)$, ii)  measurements on $\mathcal{BR}(B_{s,\,d}^0\to\mu^+\mu^-)$ and the anomalous magnetic moments of the muon $\Delta a_{\mu}$. We also analyze the upper-limit on $\mathcal{BR}(h \to \ell_i\ell_j)$ \cite{CMS:2021rsq, ATLAS:2023mvd}. However, we find it is not enough restricted. We implement the model in the \texttt{Mathematica} package so-called $\texttt{SpaceMath}$ \cite{Arroyo-Urena:2020qup} to analyze the FNSM parameter space.
	We present in Fig.~\ref{fig:RXs} the ultraviolet mass scale $\Lambda$ as a function of the VEV of the complex singlet $v_s$. The different points represent individual $\mathcal{R}_X's$; the blue circles (green triangles, yellow diamonds, green squares, orange triangles, green rectangles) correspond to $\mathcal{R}_W$ ($\mathcal{R}_{Z},\, \mathcal{R}_{\gamma},\,\mathcal{R}_{\tau},\,\mathcal{R}_{b})$. The common region of all them is represented by the enclosed area in solid black lines. The green rectangle stands for the allowed area by all the LFVp\footnote{According to our analysis, the upper bound on $\Lambda$ and $v_s$ is imposed by the anomalous magnetic moment of the muon. However, the situation could change because is still possible that more precise determinations of the SM hadronic contribution and the experimental
		measurement would settle the discrepancy in the
		future without requiring any new physics effects.}. Meanwhile, the cyan area is the result of applying all discussed theoretical and experimental constraints (intersection). Motivated by the analysis done in Sec.~\ref{theoretical_constraints}, we again scan on the model's parameters involved in the evaluation of the $\mathcal{R}_{X}$'s, as shown in Table \ref{Rs_Constraints}.  We also have explored the allowed values by $\mathcal{R}_{\mu,\,c}$ (not presented in Fig.~\ref{fig:RXs}), however, these measurements are not very stringent in the FNSM. It is worth to highlight the fact $v_s\approx\Lambda$, this is to be expected because the coupling $hf\bar{f}$ behaves as $v_s^2/\Lambda^2$ for $\cos\alpha\approx -1$, as shown in Table \ref{Diagonal_couplings}.  
	\begin{table}
		
		\caption{Scanned parameters.}\label{Rs_Constraints}
		
		\begin{centering}
			\begin{tabular}{|c|c|}
				\hline 
				Parameter & Interval\tabularnewline
				\hline 
				\hline 
				$\cos\alpha$ & {[}-1, -0.7{]}\tabularnewline
				\hline 
				$\Lambda$ & $v$-5000 GeV\tabularnewline
				\hline 
				$v_{s}$ & $v$-5000 GeV\tabularnewline
				\hline 
			\end{tabular}
			\par\end{centering}
	\end{table}

%\subsubsection{Neutral meson mixing}
%The effective Hamiltonian describing $\Delta F=2$ interactions is given by
%\begin{equation}
%	\mathcal{H}_{NP}^{\Delta F=2}=C_1^{ij}(\bar{q}_L^i\gamma_{\mu}q_L^j)^2+\tilde{C}_1^{ij}(\bar{q}^i_R \gamma_\mu q_R^j)^2+C_2^{ij}(\bar{q}_R^iq_L^j)^2
%\end{equation}
%	\begin{figure}[!hbt]
%	\subfigure[]{\includegraphics[width=7cm]{DDmixing.png}}
%	%\subfigure[]{\includegraphics[width=6.9cm]{ptb2_4b-channel.png}}
%	%\subfigure[]{\includegraphics[width=7cm]{ptb3_4b-channel.png}}
%	%\subfigure[]{\includegraphics[width=7cm]{ptb4_4b-channel.png}}
%	%\includegraphics[width=8cm]{Minv.png}
%	\caption{}\label{fig:mixings}
%\end{figure}
\subsubsection{Neutral meson mixing}

The effective
Hamiltonian describing $\Delta F =2$ interactions is given by
\begin{align}	\label{eq:heffdf2}
	\mathcal{H}_{eff}^{\Delta F=2}&=C_1^{ij} \,( \bar q^i_L\,\gamma_\mu \, q^j_L)^2+\widetilde C_1^{ij} \,( \bar q^i_R\,\gamma_\mu \, 
	q^j_R)^2 +C_2^{ij} \,( \bar q^i_R \, q^j_L)^2\\
	&+\widetilde C_2^{ij} \,( \bar q^i_L \, q^j_R)^2\notag
	+ C_4^{ij}\, ( \bar q^i_R \, q^j_L)\, ( \bar q^i_L \, q^j_R)\,+C_5^{ij}\, ( \bar q^i_L \,\gamma_\mu\, q^j_L)\, ( \bar q^i_R \,
	\gamma^\mu q^j_R)\,+ \text{h.c.} 
\end{align}
At tree level, flavon exchange leads to the generation of Wilson coefficients\cite{Buras:2013rqa,Crivellin:2013wna}
\begin{align}
	C_2^{ij} &= -(g_{ji}^*)^2\left(\frac{1}{m_s^2}-\frac{1}{m_a^2}\right)\notag \\
	\tilde C_2^{ij} &= -g_{ij}^2\left(\frac{1}{m_s^2}-\frac{1}{m_a^2}\right)\notag \\
	C_4^{ij} &= -\frac{g_{ij}g_{ji}}{2}\left(\frac{1}{m_s^2}+\frac{1}{m_a^2}\right)\,.
	\label{eq:wilsons}
\end{align}

For $M_{A_F}=M_{H_F}$ the two contributions to $C_2$ and $\widetilde C_2$
cancel, whereas constructive interference can be generated via the Wilson coefficient $C_4$.
% Given
%that the masses of the scalar and pseudo-scalar components
%\begin{align}\label{eq:masses}
%	M_{H_F} = \mu_S\,= \sqrt{\lambda_S}v_s 
%	\quad \text{and} \quad
%	M_{A_F} = \sqrt{2 b} \; 
%\end{align}
%are determined by independent scales, any occurrence of cancellation would be by coincidence.  Depending on the specific mesonic system, a significant enhancement may result from the renormalization group running, and matrix effects.

\subsubsection*{$K-\bar K$ mixing}
The limits from $K-\bar K$ mixing at 95\% C.L. \cite{Bona:2007vi} are given by
\begin{align}
	C_{\epsilon_K}&=\frac{ \text{Im} \langle K^0|\mathcal{H}^{\Delta F=2}|\bar K^0\rangle}{\text{Im} \langle K^0| \mathcal{H}_\text{SM}^{\Delta F=2} |\bar K^0 \rangle} = 1.12_{-0.25}^{+0.27}; \notag \quad 
	C_{\Delta m_K}=\frac{\text{Re}\langle K^0|\mathcal{H}^{\Delta F=2}|\bar K^0\rangle}{\text{Re} \langle K^0| \mathcal{H}_\text{SM}^{\Delta F=2} |\bar K^0 \rangle} = 0.93_{-0.42}^{+1.14} \; 
\end{align}
The contributions of SM and flavon are included in $\mathcal{H}^{\Delta F=2}$, while $\mathcal{H}_\text{SM}^{\Delta F=2}$
represents the contribution of SM. The observed dip characteristic arises from the accidental cancellation that occurs in $C_2^{sd}$ and $\tilde{C}_2^{sd}$, as shown in the Wilson coefficients generated by flavon exchage at tree-level ~\cite{Buras:2013rqa,Crivellin:2013wna}. This feature is universally observed within in $K-\bar{K}$ mixing. However, it is absent in the specific scenarios where the contribution to $C_4^{sd}$ becomes so dominant that it overshadows all other contributions.

\subsubsection*{$B-\bar B$ mixing}
For both variations of $B -\bar B$ mixing, it is defined as follows,
\begin{align}
	C_{B_q}e^{2i\varphi_{B_q}}=\frac{\langle B_q|\mathcal{H}^{\Delta F=2}|\bar B_q\rangle}{\langle B_q| \mathcal{H}_\text{SM}^{\Delta F=2} |\bar B_q \rangle}\; ,
\end{align}
with the 95\%~CL limits~\cite{Bona:2007vi} given by

\begin{align}
	C_{B_d}&= 1.05\pm 0.11,  & \varphi_{B_d}&= -2.0\pm 1.8, \notag \\
	C_{B_s}&= 1.110\pm 0.090, & \varphi_{B_s}&= 0.42\pm 0.89 \; .
\end{align}
\subsubsection*{$D-\bar D$ mixing}

\begin{figure}[t]
	\begin{center}
		\includegraphics[width=.42\textwidth]{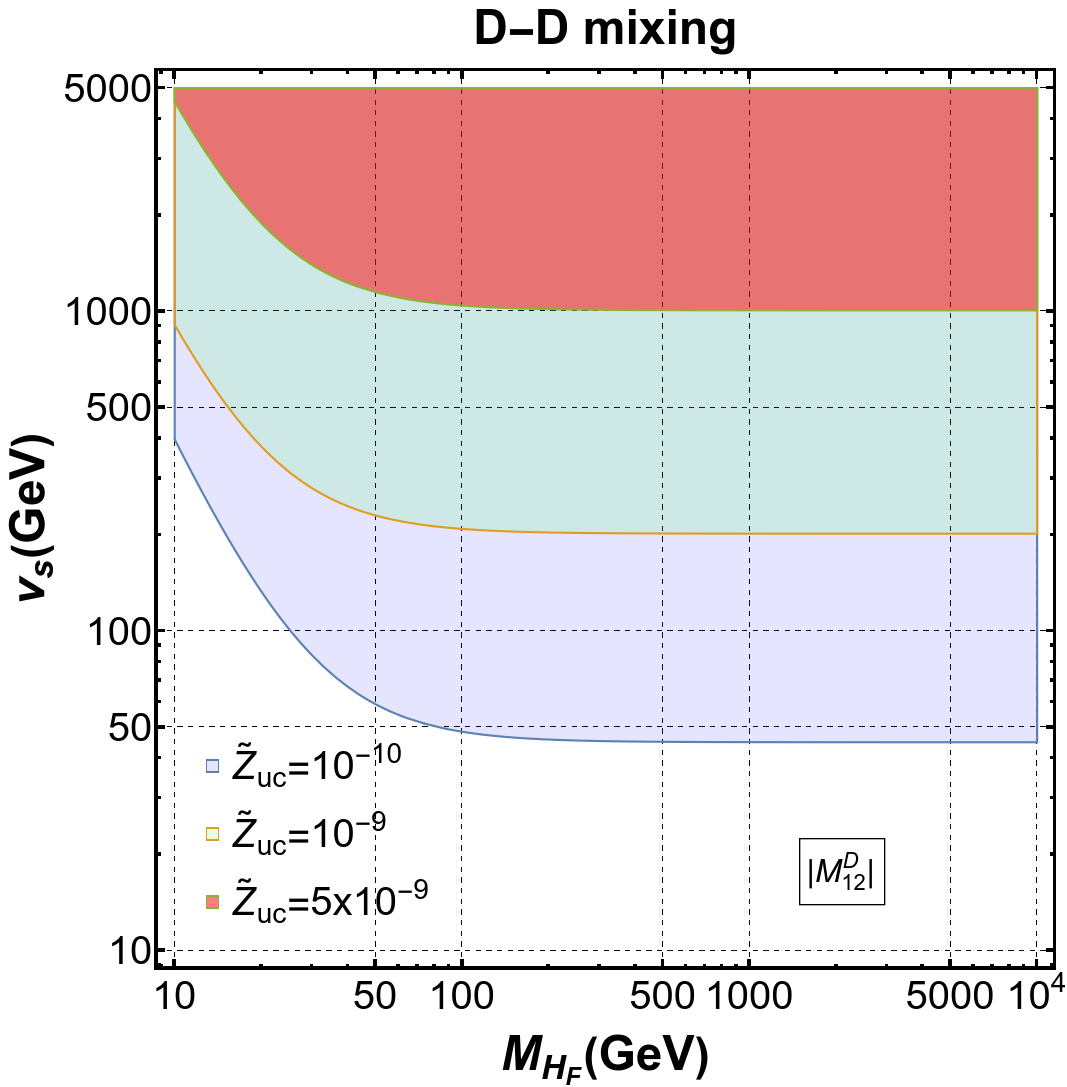}
	\end{center}
	\caption{$M_{H_F}-v_s$ plane showing the region allowed due to flavon contributions to $|M_{12}^D|$. Blue area: $\tilde{Z}_{uc}$=$10^{-10}$, green area: $\tilde{Z}_{uc}$=$10^{-9}$, red area: $\tilde{Z}_{uc}$=$5\times 10^{-9}$. }
	\label{fig:DDmix}
\end{figure}

Given that the SM contribution to the $D-\bar D$ mixing process is significantly affected by substantial hadronic uncertainties, it is recommended to define and restrict the contributions of flavons ensuring that they do not exceed the constraint $2\sigma$\cite{Bevan:2014tha}.
\begin{align}
	|M^{D}_{12}|
	=|\langle D|\mathcal{H}^{\Delta F=2}|\bar D\rangle
	< 7.5\times 10^{-3}~\text{ps}^{-1} \; .
\end{align}
In Refs. \cite{Bauer:2016rxs, Abbas:2024dfh} are presented stringent restrictions in the $v_s-M_{A_F}$ plane that come from mixtures of mesons. However, we evade them by inheriting them to the parameters $\tilde{Z}_{ds}$, $\tilde{Z}_{db}$ and $\tilde{Z}_{uc}$ for $K-\bar{K}$, $B-\bar{B}$ and $D-\bar{D}$, respectively. We presented in Fig. \ref{fig:DDmix} the $M_{H_F}-v_s$ plane for three different values of $\tilde{Z}_{uc}$; $10^{-10}$, $10^{-9}$ and $5 \times 10^{-9}$. The colored areas are these allowed by $|M_{12}^D|$. Similar constraints are obtained for $K-\bar{K}$ and $B-\bar{B}$. 

	%%%%%%%%%%%%%%%%%%%%%%%%%%%%%%%%%%%%%%%%%%%%%%%%%%%%%%%%%%%%%%%%%%%%%%%%%%%%%%%%%%%%%%%%%%%%%%%%%%%%%%%%%%%%%%%%%%%%%%%%%%%%%%%%%%%%%%%%%%%%%%%%%%%%%%%%%%%%%%%%%%%%%%%%%%%%%%%%%%%%%%%%%%%%%%%%%%%%%

	\section{Collider analysis }
	\label{se:col_an}
	
	We first present the decay width $\Gamma(H_F\to h\bar{f}f)$ of the three-body process in which we are interested. The study of these kinds of processes is interesting as it can also have a sizable branching ratio. The Feynman diagrams that contribute to these reactions are shown in Fig. \ref{Htoffh}. 
	
	The decay width can be written as follows
	\begin{eqnarray}
		\Gamma(H_F\to h\bar{f}f)&=&\frac{M_{H_F}}{256 \pi
			^3} \int  dx_a\int dx_b |\overline {\cal M} |^2,
	\end{eqnarray}
	where the average square amplitude is given by
	\begin{align}
		|\overline {\cal M}|^2
		&= \frac{1}{2\left(x_a+x_b+x_h-2\right)^2}\left(x_a+x_b+x_h-4	x_t-1\right)\left(
		\left(x_a+x_b+x_h-2\right)C_a+C_b\right)^2\nonumber\\
		&+\frac{2 }{\left(x_a-1\right)^2\left(x_b-1\right)^2}\Bigg(\left(x_a-1\right)\left(x_b-1\right)
		\left(x_a-x_b\right)^2-16\left(x_a+x_b-2\right)^2 x_t^2\nonumber\\
		&+4	\left(x_a+x_b-2\right) \left(2-3x_b+x_a \left(4 x_b-3\right)\right)x_t+x_h \left(4
		\left(x_a+x_b-2\right)^2x_t-\left(x_a-x_b\right)^2\right)\Bigg)C_c^2\nonumber\\
		&-\frac{4 \sqrt{x_t}}{\left(x_a-1\right)\left(x_b-1\right)}	(x_a^2+2 \left(3 x_b+x_h-4
		x_t-3\right) x_a+x_b^2-4 x_h+2 x_b\left(x_h-4 x_t-3\right)\nonumber\\
		&+16x_t+4)C_a C_c-\frac{4 \sqrt{x_t}}{\left(x_a-1\right)\left(x_b-1\right)\left(x_a+x_b+x_h-2\right)} \Bigg(x_a^2+2\left(3 x_b+x_h-4 x_t-3\right)x_a\nonumber\\
		&+x_b^2-4 x_h+2 x_b \left(x_h-4	x_t-3\right)+16	x_t+4\Bigg)C_b C_c.
	\end{align}
	with $x_a=(m_a/M_{H_F})^2$. The factors $C_a=g_{H_F hff}$, $C_b=g_{H_F h h}g_{h ff}/m_h^2$, and $C_c=g_{H_F ff}g_{hff}/m_h$ are the coupling constants involved in the Feynman diagrams of Fig. \ref{Htoffh}.
	%\begin{figure}[!hbt]
	%	\includegraphics[width=10cm]{Htohff.pdf}
	%	\caption{Feynman diagrams inducing the $H\to \bar{f}fh$ decay in the FNSM.\label{Htoffh}}
	%\end{figure}
	Finally, the integration domain is given by
	\begin{equation}
		2 \sqrt{x_t}\leq x_a\leq 1-x_h-2 \sqrt{x_t x_h},
	\end{equation}
	\begin{equation}
		x_b \gtreqqless\frac{2 (1-x_h+2x_t)+x_a \left(x_a+x_h-2
			x_t-3\right)\mp\sqrt{x_a^2-4 x_t}
			\sqrt{\left(x_a+x_h-1\right)^2-4 x_h
				x_t}}{2 \left(1-x_a+x_t,
			\right)}
	\end{equation}
	
	Meanwhile, the production cross-section of the heavy CP-even Flavon  $H_F$ (or pseudo scalar $A_F$, for that matter) depends mainly on the $g_{H_F t\bar{t}}= \frac{ c_\alpha v + s_\alpha
		v_s}{v_s}\, \frac{y_t}{\sqrt{2}}$ ($g_{A_F t\bar{t}}=\frac{v}{v_s}\, \frac{y_t}{\sqrt{2}} $) coupling.  
	The corresponding term in the effective Lagrangian reads \cite{Plehn:2009nd}:
	%%%%%%%%%%%
	%%%%%%%%%%%
	%%%%%%%%%%%
	\begin{eqnarray}
		\mathcal{L}_{\rm eff}&=&\frac{1}{v} \, g_{hgg} \, h \, G_{\mu\nu} G^{\mu\nu},\\
		g_{Sgg} &=& -i \, \frac{\alpha_S}{8\pi}\, \tau (1+(1-\tau)\,f(\tau))~~~~~{\rm with}~~\tau = \frac{4 M_t^2}{M_h^2},\\
		f(\tau)&=& \begin{cases} 
			(\sin^{-1}\sqrt{ \frac{1}{\tau} })^2, \quad\quad\quad\quad\quad\quad \tau\geq 1,\\ 
			-\frac{1}{4}[\ln\frac{1+\sqrt{ 1-\tau}}{1-\sqrt{1-\tau}}-i\pi]^2\quad\quad\quad \tau<1.
		\end{cases}
	\end{eqnarray}
	%%%%%%%%%%%
	%%%%%%%%%%%
	%%%%%%%%%%%
	In FNSM, the $ggh$, $gg H_{F}$ and $gg A_{F}$ couplings are given, respectively, by:
	\begin{eqnarray}
		g_{hgg}&=&\left(\frac{c_\alpha v_s - s_\alpha v }{v_s}\right)\, g_{Sgg},\nonumber\\
		g_{H_F gg}&=&\left (\frac{c_\alpha v +s_\alpha v_s }{v_s}\right)\, g_{S gg},\nonumber \\
		g_{A_F gg}&=&\frac{v}{v_s}\,(-i \,\alpha_S/\pi)\, \tau \,f(\tau),
	\end{eqnarray}
	$A_F$ is a CP-odd scalar and $h, H_F$ are CP-even scalars so, once the couplings with left and right fields are written in terms of Dirac fields, the Hermitian part of the coupling in Eq.~\ref{eq:fermlag} gives rise to an $i=\sqrt{-1}$ coupling for $h, H_F$ and a $\gamma_5$ coupling for $A_F$: so the result of the top quark loop integral is different for $h,H_F$ and $A_F$ \cite{Pak:2011hs,LHCHiggsCrossSectionWorkingGroup:2016ypw}. It is {to be} noted that, for $M_{H_F, A_F}>2\, M_t$, $f(\tau)=-\frac{1}{4}[\ln\frac{1+\sqrt{ 1-\tau}}{1-\sqrt{1-\tau}}-i\pi]^2$.
	
	As far as our computation scheme is concerned, we first use {\tt FeynRules} \cite{Alloul:2013bka} to obtain the FNSM model and produce the UFO files for {\tt MadGraph-2.6.5} \cite{Alwall:2014hca}. Using the ensuing particle spectrum in {\tt MadGraph-2.6.5}, we calculate the production cross-section of the aforementioned production and decay process.
	The \texttt{$\rm MadGraph\_aMC@NLO$} \cite{Alwall:2014hca} framework has been used to generate the background events in the SM. The showering and hadronization simulations were performed with \texttt{Pythia-8} \cite{Sjostrand:2014zea}.
	The detector response has been emulated using \texttt{Delphes-3.4.2}  \cite{deFavereau:2013fsa}. The default ATLAS card which comes along with the \texttt{Delphes-3.4.2} package was considered in this analysis. For both the signal and background processes, we consider the Leading Order (LO) cross-sections computed by \texttt{$\rm MadGraph\_aMC@NLO$}.

	Once we have put on the table the way in which we evaluate the collider observables, we define in Table \ref{Scenarios} three scenarios to be studied in the following analysis.
	\begin{table}[!htb]
		
		\caption{Scenarios (\textbf{S1}, \textbf{S2}, \textbf{S3}) used in the calculations.}\label{Scenarios}
		
		\begin{centering}
			\begin{tabular}{|c|c|c|c|}
				\hline 
				Parameter & \textbf{S1} & \textbf{S2} & \textbf{S3}\tabularnewline
				\hline 
				\hline 
				$\cos\alpha$ & $0.995$ & $0.995$ & $0.995$\tabularnewline
				\hline 
				$\Lambda$ & $1$ TeV & $1.5$ TeV & $2.5$ TeV\tabularnewline
				\hline 
				$v_{s}$ & $1$ TeV & $1.5$ TeV & $2.5$ TeV\tabularnewline
				\hline 
			\end{tabular}
			\par\end{centering}
	\end{table}
	
	We present in Table \ref{XS-signal} the numerical cross-section of the proposed signal and the number of events produced by considering scenarios $\textbf{S1},\,\textbf{S2},\,\textbf{S3}$.
	\begin{table}[!htb]
		
		\caption{Cross-section of the signal for scenarios $\textbf{S1},\,\textbf{S2},\,\textbf{S3}$.}\label{XS-signal}
		
		\begin{centering}
			\begin{tabular}{|c|c|c|c|}
				\hline 
				Scenario & $M_{H_{F}}$(GeV) & $\sigma(pp\to H_{F}\to hb\bar{b})$(fb) & Events $(\mathcal{L}_{{\rm int}}=300$fb$^{-1})$\tabularnewline
				\hline 
				\hline 
				$\textbf{S1}$ & $(800,\,900,\,1000)$ & $(6.8,\,3.3,\,1.7)$ & $(2040,\,990,\,510)$\tabularnewline
				\hline 
				$\textbf{S2}$ & $(800,\,900,\,1000)$ & $(4.3,\,2.3,\,1.3)$ & $(1290,\,690,\,390)$\tabularnewline
				\hline 
				$\textbf{S3}$ & $(800,\,900,\,1000)$ & $(3.8,\,2.1,\,1.1)$ & $(1140,\,630,\,330)$\tabularnewline
				\hline 
			\end{tabular}
			\par\end{centering}
	\end{table}
	
\begin{figure}[!htb]
		\begin{center}
			\includegraphics[scale=0.25]{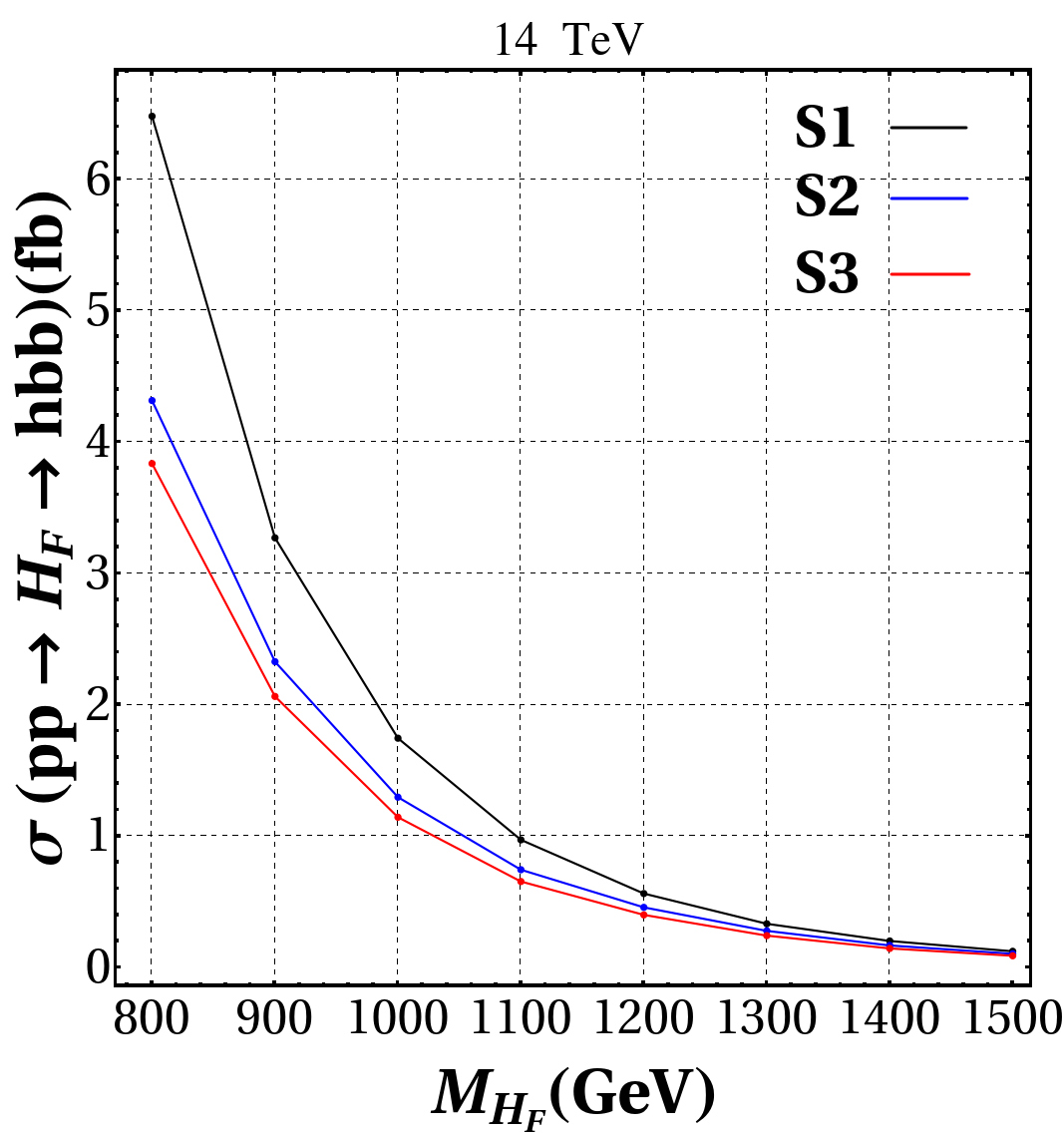}
		\end{center}
		\caption{Production cross-section of the signal $pp\to H_F\to h b\bar{b}$. The centre-of-mass energy was set to 14 TeV.}
		\label{XS1}
	\end{figure}

Meanwhile, Fig. \ref{XS1} presents an overview of the production cross-section of the signal $pp\to H_F\to h b\bar{b}$ by considering the three scenarios $\textbf{S1},\,\textbf{S2},\,\textbf{S3}$.
%	We observe that scenario $\textbf{S1}$ ($\textbf{S2},\,\textbf{S3}$), for $M_{H_F}=800$ GeV and an integrated luminosity ($\mathcal{L}_{\rm{int}}$) of $300$ fb$^{-1}$, predicts $1935$ ($1290,\,1140$) signal events ($hb\bar{b}$).   
	
	On the other hand, we analyze two particular channels in which the Higgs boson decays, namely,  i) $pp \to H_F\to h b \bar{b}(h\to b \bar{b})$ and ii) $pp \to H_F\to h b \bar{b}(h\to \gamma \gamma)$.
	The $b$-tagging of jets produced from the fragmentation and hadronization of bottom quarks represents a fundamental role in separating the signal from the background processes, which involve gluons, light-flavor jets ($u,\,d,\,s$) and c-quark fragmentation. To overcome this problem, we use the \texttt{FastJet} package~\cite{Cacciari:2011ma} (via \texttt{MadAnalysis}~\cite{Conte:2012fm}) and invoke the anti-$k_T$  algorithm~\cite{Cacciari:2008gp}. We also include the $b$-tagging efficiency $\epsilon_b=90\%$. The probability that a $c-$jet or any other light-jet $j$ is mistagged as a $b-$jet are $\epsilon_c=5\%$ and $\epsilon_j=1\%$~\cite{ATLAS:2023gog}, respectively.

	\subsubsection{$pp \to H_F\to h b \bar{b}(h\to b \bar{b})$}
	The SM background processes for the $h\to b \bar{b}$ channel are given by
	\begin{itemize}
		\item $pp\to t\bar{t},\,(t\to W^+ \bar{b},\,W^+\to c\bar{b},\,\bar{t}\to W^- b,\,W^+\to b\bar{c}$),
		\item $pp\to Wh,\, (W\to cb,\, h\to b\bar{b}$),
		\item $pp\to ZZ,\, (Z\to b\bar{b},\, Z\to b\bar{b}$),
		\item $pp\to Zh,\, (Z\to b\bar{b},\, h\to b\bar{b}$),
		\item $pp\to b\bar{b}jj$, where $j$ denotes non-bottom-quark jets. 
	\end{itemize} 
The numerical cross-section of the SM background processes is presented in Table~\ref{XSbgd_4b}.
	
	\begin{table}
		
		\caption{Cross-section of the SM backgrounds processes.}\label{XSbgd_4b}
		
		\begin{centering}
			\begin{tabular}{cc}
				\hline 
				SM background process & Cross-section (fb)\tabularnewline
				\hline 
				\hline 
				$pp\to t\bar{t}$ & $26910$\tabularnewline
				\hline 
				$pp\to Wh$ & $0.5463$\tabularnewline
				\hline 
				$pp\to ZZ$ & $231.5$\tabularnewline
				\hline 
				$pp\to Zh$ & $79.75$\tabularnewline
				\hline 
				$pp\to\bar{b}bjj$ & $5.441\times10^{8}$\tabularnewline
				\hline 
			\end{tabular}
			\par\end{centering}
	\end{table}

	For this channel, the $b$-jets emerging from the primary vertex are expected to have a high transverse momentum $p_T(b_1,\,b_2)$, while these produced via the decay $h\to b\bar{b}$ have a lower transverse momentum than the primary ones $p_T(b_3,\,b_4)$. An important fact, and the clearest signature of our signal, is the resonant effect coming from the decay $H_F\to h b\bar{b}\,(h\to b\bar{b})\to b\bar{b}b\bar{b}$. 
	
\begin{figure}[!hbt]
		\includegraphics[width=8cm]{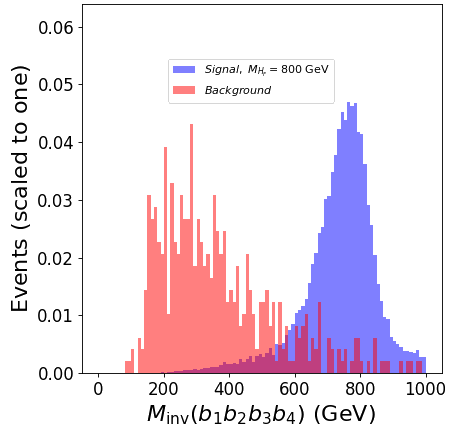}
		\caption{Normalized distribution of the reconstructed invariant mass $M_{\rm inv}(b_1b_2b_3b_4)$ for the signal and background processes.}\label{Minv4b}
	\end{figure}

	We present in Fig. \ref{Minv4b} the invariant mass $M_{\rm inv}(b_1b_2b_3b_4)$ for scenario $S1$ and $M_{H_F}=800$ GeV. 
	
\begin{figure}[!hbt]
		\includegraphics[width=7cm]{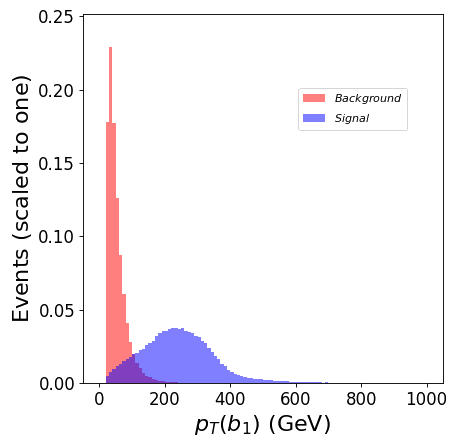}
		\includegraphics[width=6.9cm]{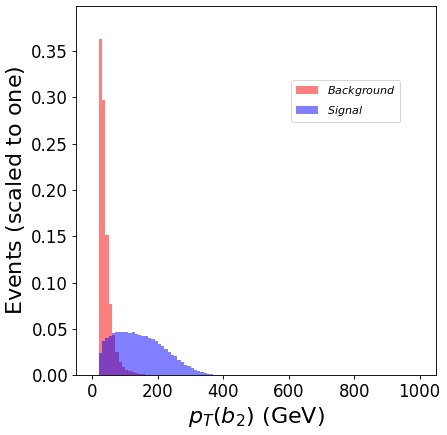}
		\includegraphics[width=7cm]{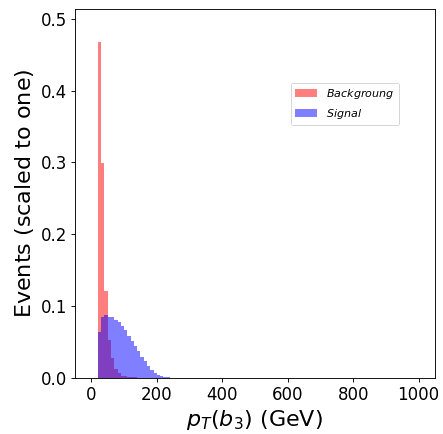}
		\includegraphics[width=7cm]{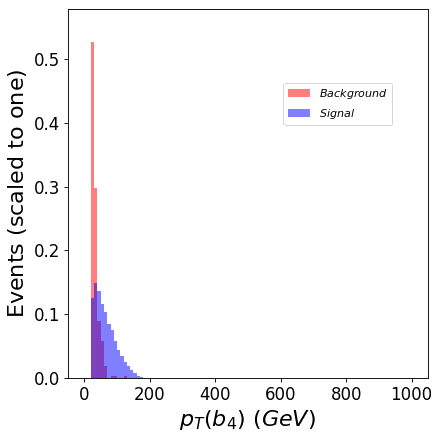}
		\caption{Normalized distributions in $b$-jet transverse momentum for signal and total background after the acceptance cuts.}\label{distributions_bb-channel}
	\end{figure}

	Meanwhile, Fig. \ref{distributions_bb-channel} shows the $p_T(b_i)$ ($i=1,\,2,\,3,\,4$) distribution including to the signal and the SM background processes. Subscript $1(4)$ corresponds to the $b$-jet with the highest (lowest) transverse momentum. While subscript 2(3) represents the second (third) dominant $b$-jet.
	
	From Fig. \ref{distributions_bb-channel} we note that $p_T(b_1,\,b_2)$ are higher than $p_T(b_3,\,b_4)$ because the former come from the primary vertex, while the last arise from the Higgs boson decay. Both $p_T(b_i)$ and $M_{\rm inv}(b_1b_2b_3b_4)$ are the most important variables to isolate the siganl from the background.
	
	\subsubsection{$pp \to H_F\to h b \bar{b}(h\to \gamma \gamma)$}
	As far as the di-photon channel is concerned, the SM background processes are given by
	\begin{itemize}
		\item $pp\to h t\bar{t},\,(h\to\gamma\gamma,\,t\to W^+ \bar{b},\,W^+\to \ell^+ \nu_\ell,\,\bar{t}\to W^- b,\,W^+\to \ell^- \bar{\nu}_\ell$),
		\item $pp\to t\bar{t}\gamma\gamma,\,(t\to W^+ \bar{b},\,W^+\to \ell^+ \nu_\ell,\,\bar{t}\to W^- b,\,W^+\to \ell^- \bar{\nu}_\ell$),
		\item $pp\to Wh,\, (W\to cb,\, h\to \gamma\gamma$),
		\item $pp\to Zh,\, (Z\to b\bar{b},\, h\to \gamma\gamma$),
		\item $pp\to hjj,\,  (h\to \gamma\gamma$),
		\item $pp\to \gamma\gamma jj$
		\item $pp\to \gamma\gamma b\bar{b}$.
	\end{itemize} 
The numerical cross-section of the SM background processes is presented in Table~\ref{XSbgd_2b2g}.
\begin{table}
		\caption{Cross-section of the SM background processes.}\label{XSbgd_2b2g}
		\begin{centering}
		\begin{tabular}{cc}
			\hline 
			SM background processes & Cross-section (fb)\tabularnewline
			\hline 
			\hline 
			$pp\to ht\bar{t}$ & $3.2\times10^{-2}$\tabularnewline
			\hline 
			$pp\to t\bar{t}\gamma\gamma$ & $0.57$\tabularnewline
			\hline 
			$pp\to Wh$ & $9.7\times10^{-4}$\tabularnewline
			\hline 
			$pp\to Zh$ & $0.14$\tabularnewline
			\hline 
			$pp\to hjj$ & $13.62$\tabularnewline
			\hline 
			$pp\to\gamma\gamma jj$ & $1.1\times10^{5}$\tabularnewline
			\hline 
			$pp\to b\bar{b}\gamma\gamma$ & $5113$\tabularnewline
			\hline 
		\end{tabular}
		\par\end{centering}
\end{table}

\begin{figure}[!hbt]
	\includegraphics[width=8cm]{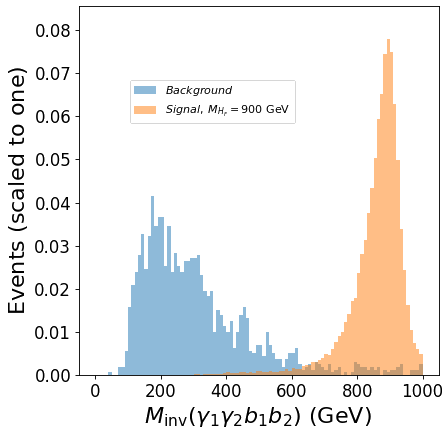}
	\caption{Normalized distribution of the reconstructed invariant mass $M_{\rm inv}(\gamma_1\gamma_2 b_1b_2)$ for the signal and background processes.}\label{Minv2b2gam}
\end{figure}

\begin{figure}[!hbt]
		\includegraphics[width=7cm]{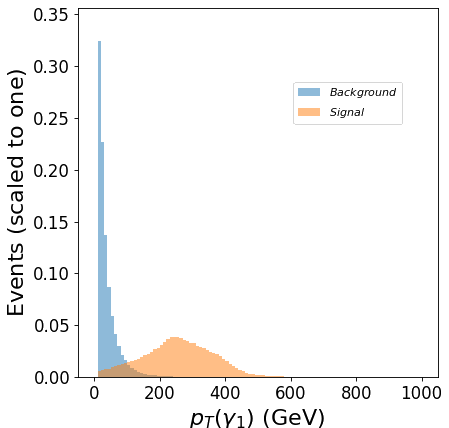}
		\includegraphics[width=6.9cm]{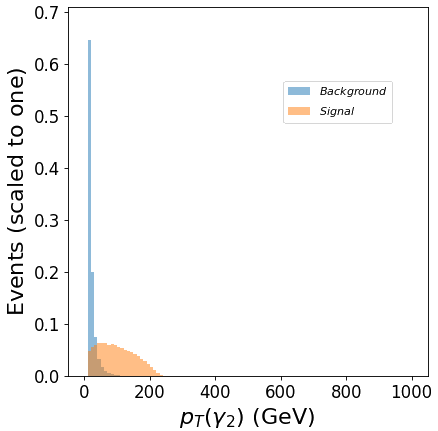}
		\includegraphics[width=7cm]{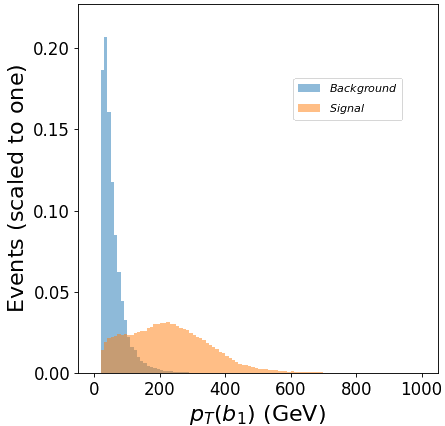}
		\includegraphics[width=7cm]{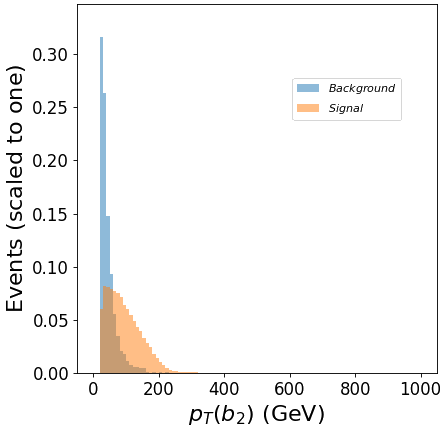}
		\caption{Normalized distributions in $b$-jet and $\gamma$ transverse momentum for signal and total background after the acceptance cuts.}\label{distributions_gammagamma-channel}
	\end{figure}

In this case we also have a similar scenario as in the previous channel concerning to the resonant effect coming from the decay $H_F\to b\bar{b}\gamma\gamma$. Figure \ref{Minv2b2gam} displays the invariant mass distribution $M_{\rm inv}(bb\gamma\gamma)$ for $M_{H_F}=900$ GeV.
	As in the previous channel, we also present in Fig. \ref{distributions_gammagamma-channel} the $p_T(b_i)$, $p_T(\gamma_i)$ ($i=1,\,2$).
	%\subsubsection{$pp \to H_F\to h b \bar{b}(h\to W W^*)$}
	
	%\subsubsection{$pp \to H_F\to h b \bar{b}(h\to Z Z^*)$}
	
	%\subsubsection{$pp \to H_F\to h b \bar{b}(h\to \tau^-\tau^+)$}
	
	\subsection{Multivariate Analysis}
		After the kinematic analysis, we found that most of the observables used to distinguish signal from background have relatively weak discriminating power. Therefore, the final candidate selection is determined using Multivariate Analysis (MVA) discriminators, which combine these observables into a single, more powerful classifier. For the MVA training, we use a boosted decision tree (BDT) algorithm \cite{Book:bdt}, implemented via the XGBoost library \cite{Chen:2016:XST:2939672.2939785}, which employs an advanced gradient boosting technique. 
%The BDT classifiers are trained using variables related to the kinematics of final and intermediate state particles, including the transverse momentum ($p_T$) and rapidity of the daughter $b$-jets (or photons, in the case $h\to \gamma \gamma$), as well as the $p_T$ of Flavon and Higgs particles. 
The BDT classifiers are trained using variables related to the kinematics of final and intermediate state particles, including:\\
Case 1: Four $b$-jets $(b_1, b_2, b_3, b_4)$:
\begin{itemize}
    \item Transverse momentum ($p_T$):  $p_T(b_1)$, $p_T(b_2)$, $p_T(b_3)$, $p_T(b_4)$
    \item Rapidity ($\eta$): $\eta(b_1)$, $\eta(b_2)$, $\eta(b_3)$, $\eta(b_4)$
    \item Combined transverse momentum: $p_T(b_1b_2b_3b_4)$
\end{itemize}
Case 2: Two $b$-jets $(b_1, b_2)$ and two photons ($\gamma_1$, $\gamma_2$):
\begin{itemize}
    \item Transverse momentum ($p_T$): $p_T(b_1)$, $p_T(b_2)$, $p_T(\gamma_1)$, $p_T(\gamma_2)$ 
    \item Rapidity ($\eta$):  $\eta(b_1)$, $\eta(b_2)$, $\eta(\gamma_1)$, $\eta(\gamma_2)$ 
    \item Invariant mass of the photons: $M(\gamma_1 \gamma_2)$ 
    \item Combined transverse momentum: $p_T(b_1b_2b_3b_4)$
\end{itemize}

\begin{figure}[!htb]
	\subfigure[]{\includegraphics[width=0.32\textwidth]{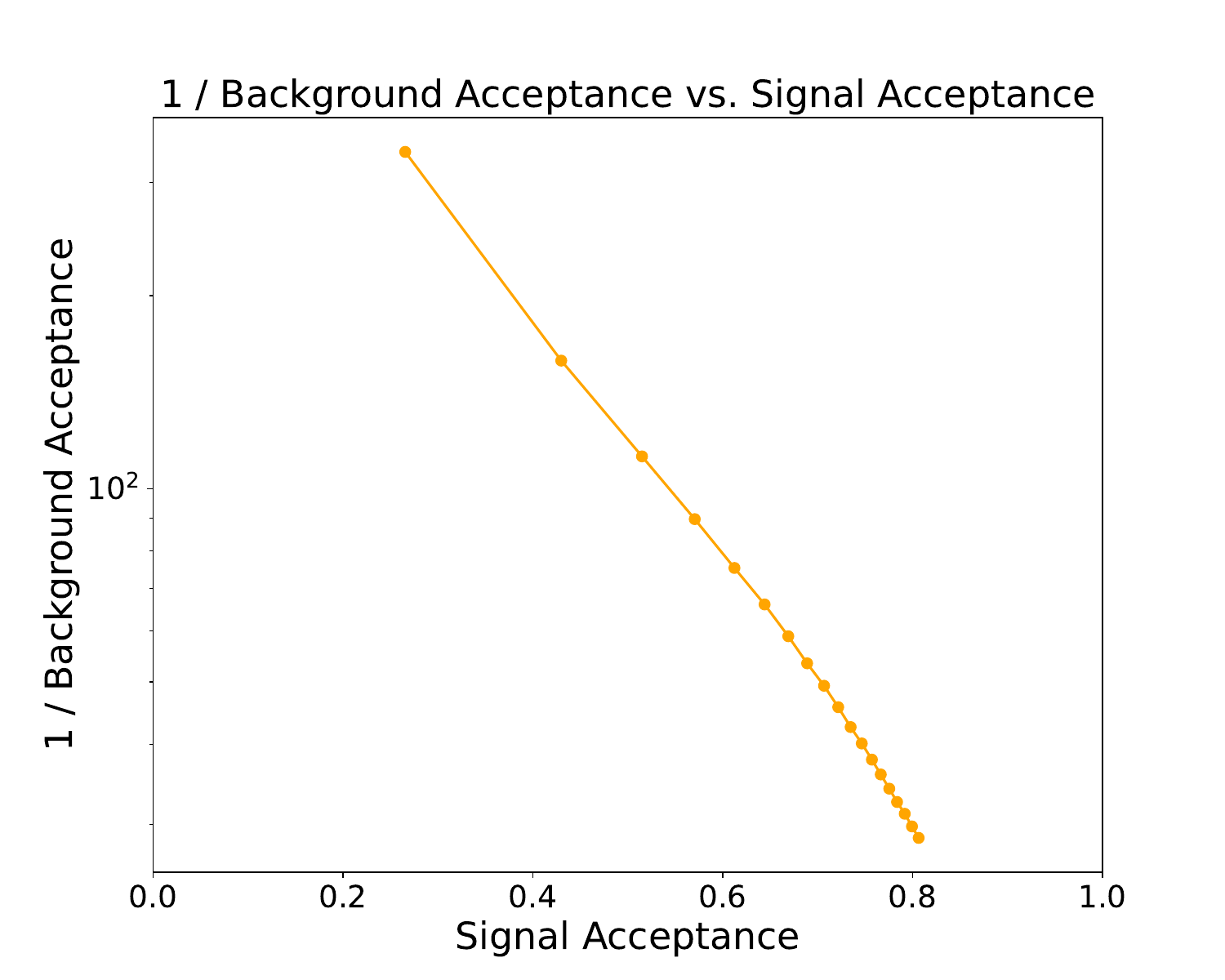}}
	\subfigure[]{\includegraphics[width=0.32\textwidth]{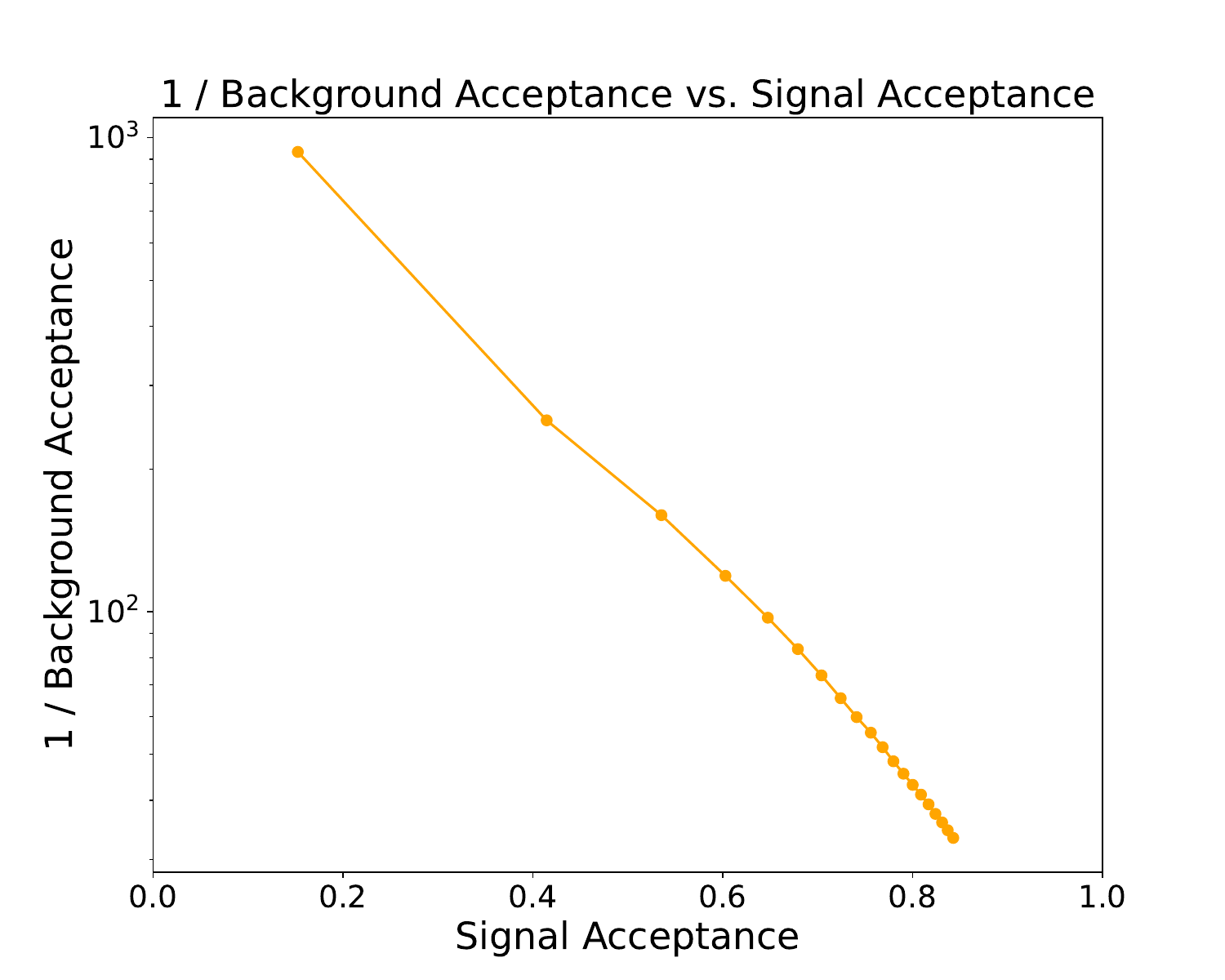}}
	\subfigure[]{\includegraphics[width=0.32\textwidth]{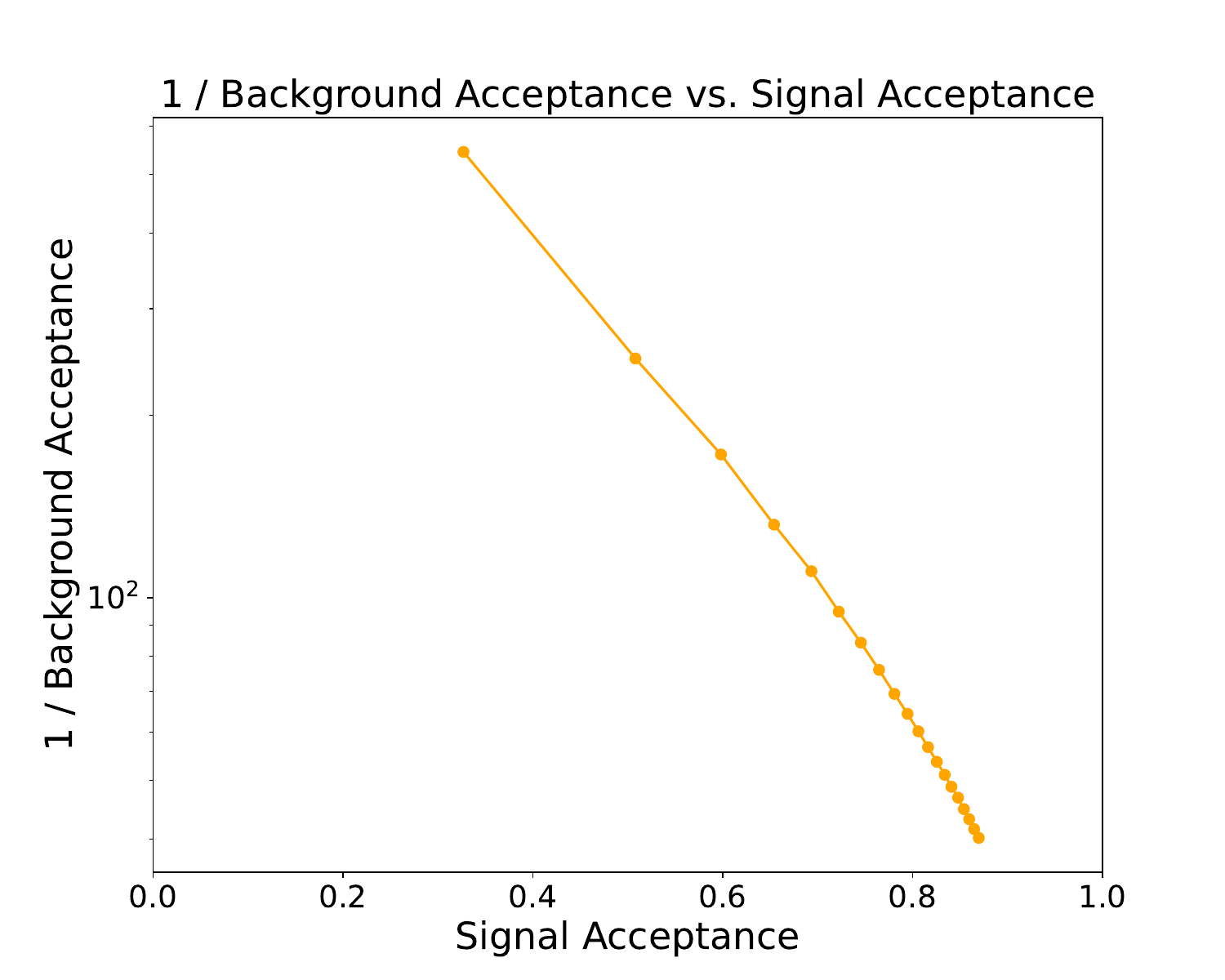}}
	\subfigure[]{\includegraphics[width=0.32\textwidth]{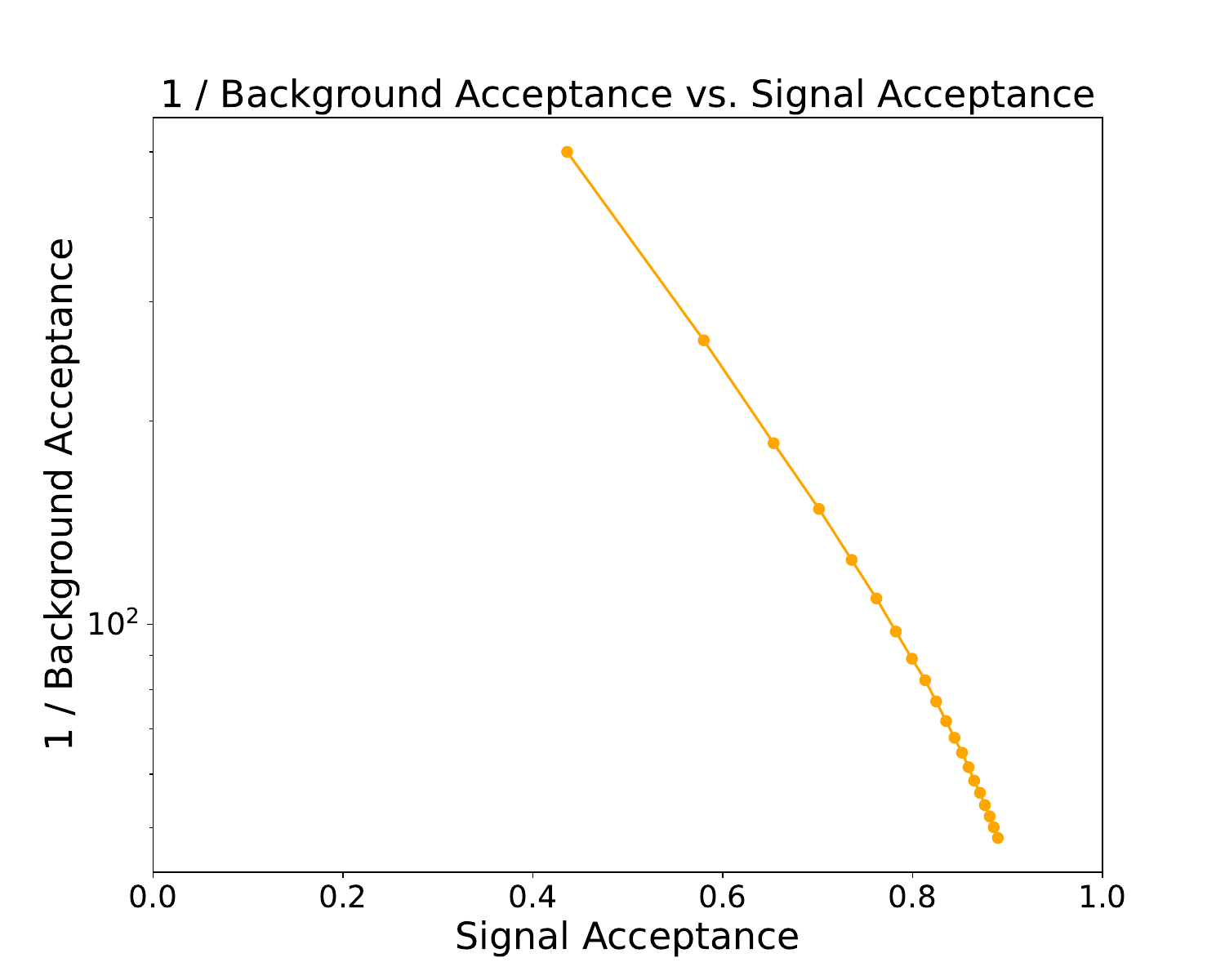}}
	\subfigure[]{\includegraphics[width=0.32\textwidth]{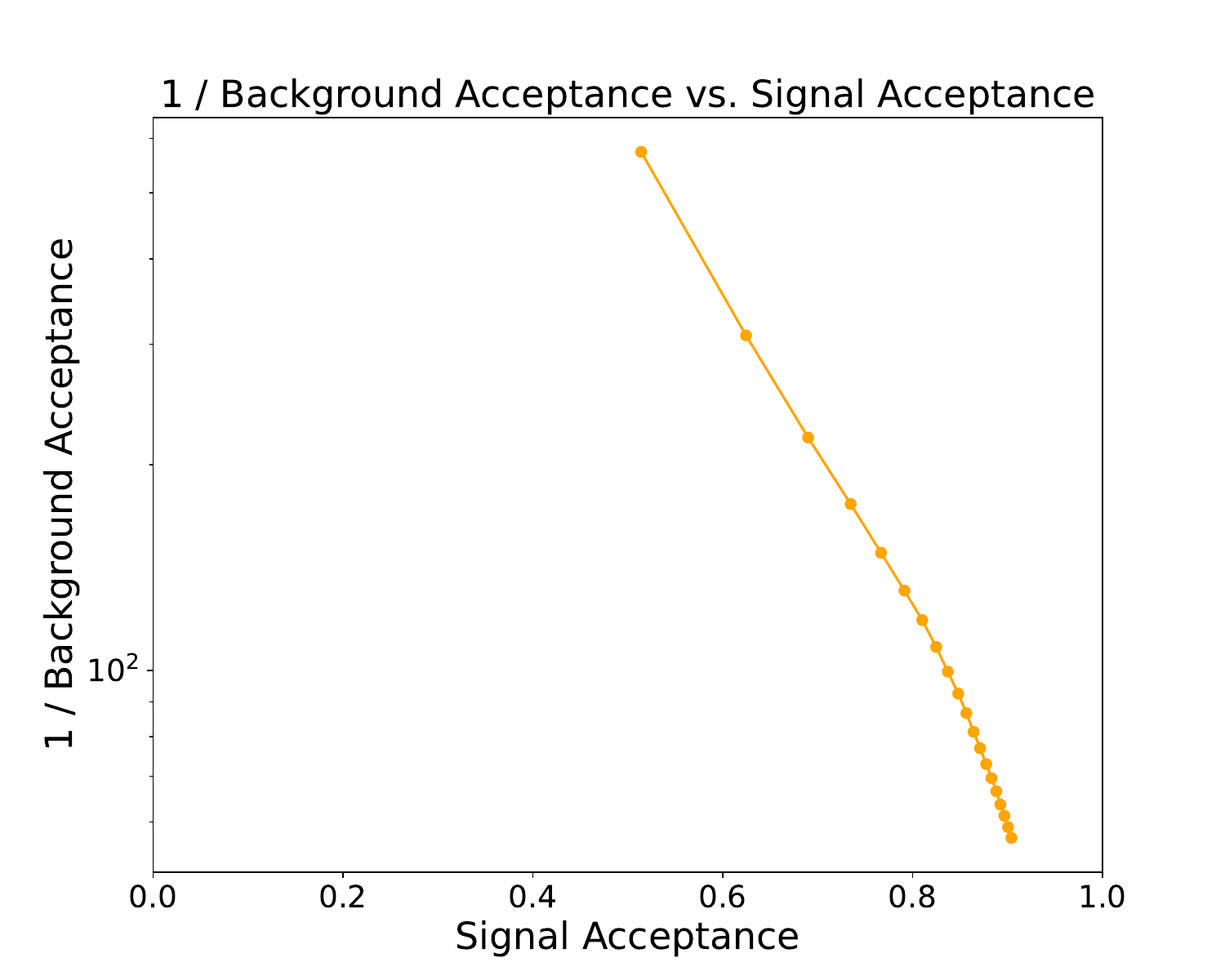}}
	\subfigure[]{\includegraphics[width=0.32\textwidth]{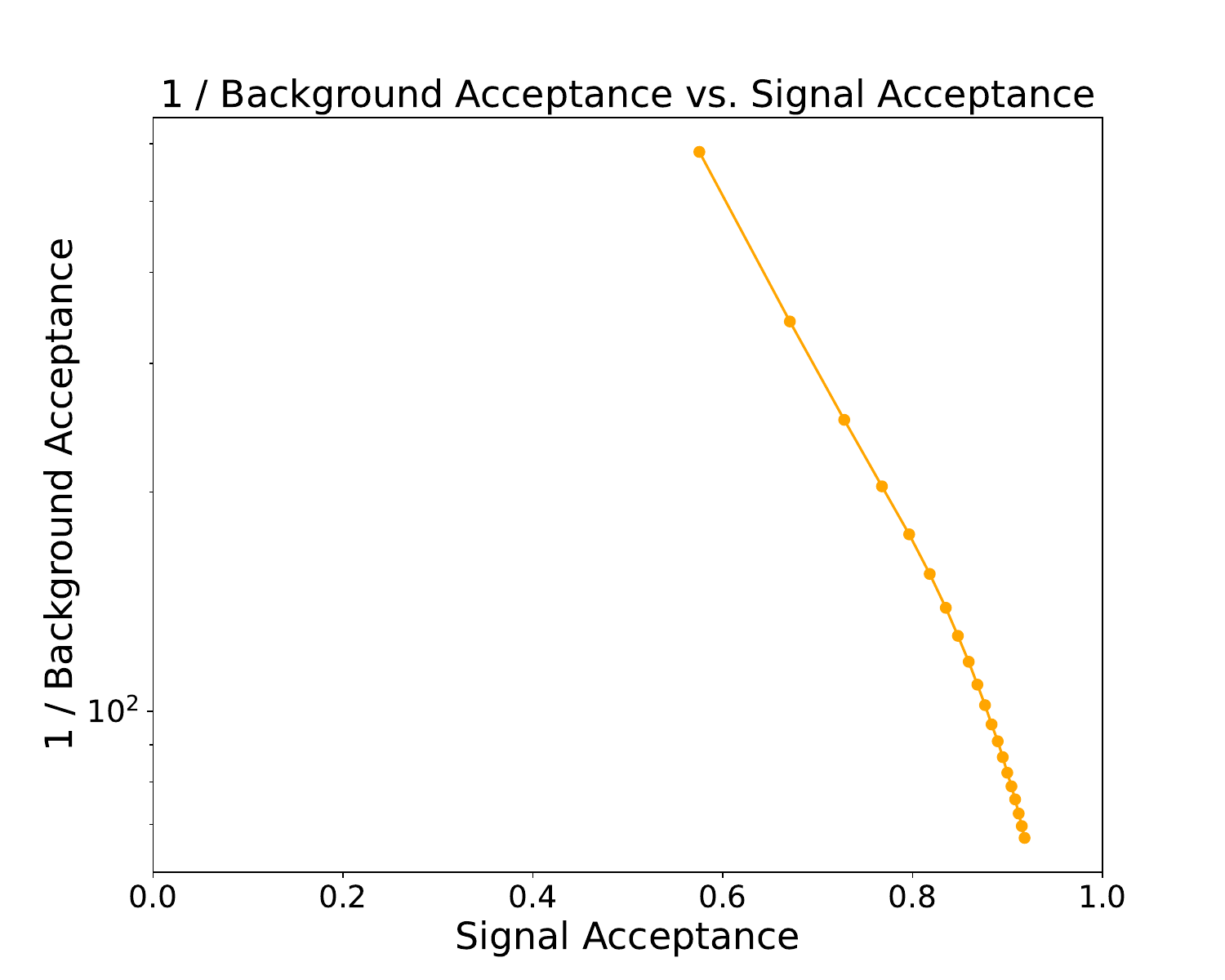}}
	\subfigure[]{\includegraphics[width=0.32\textwidth]{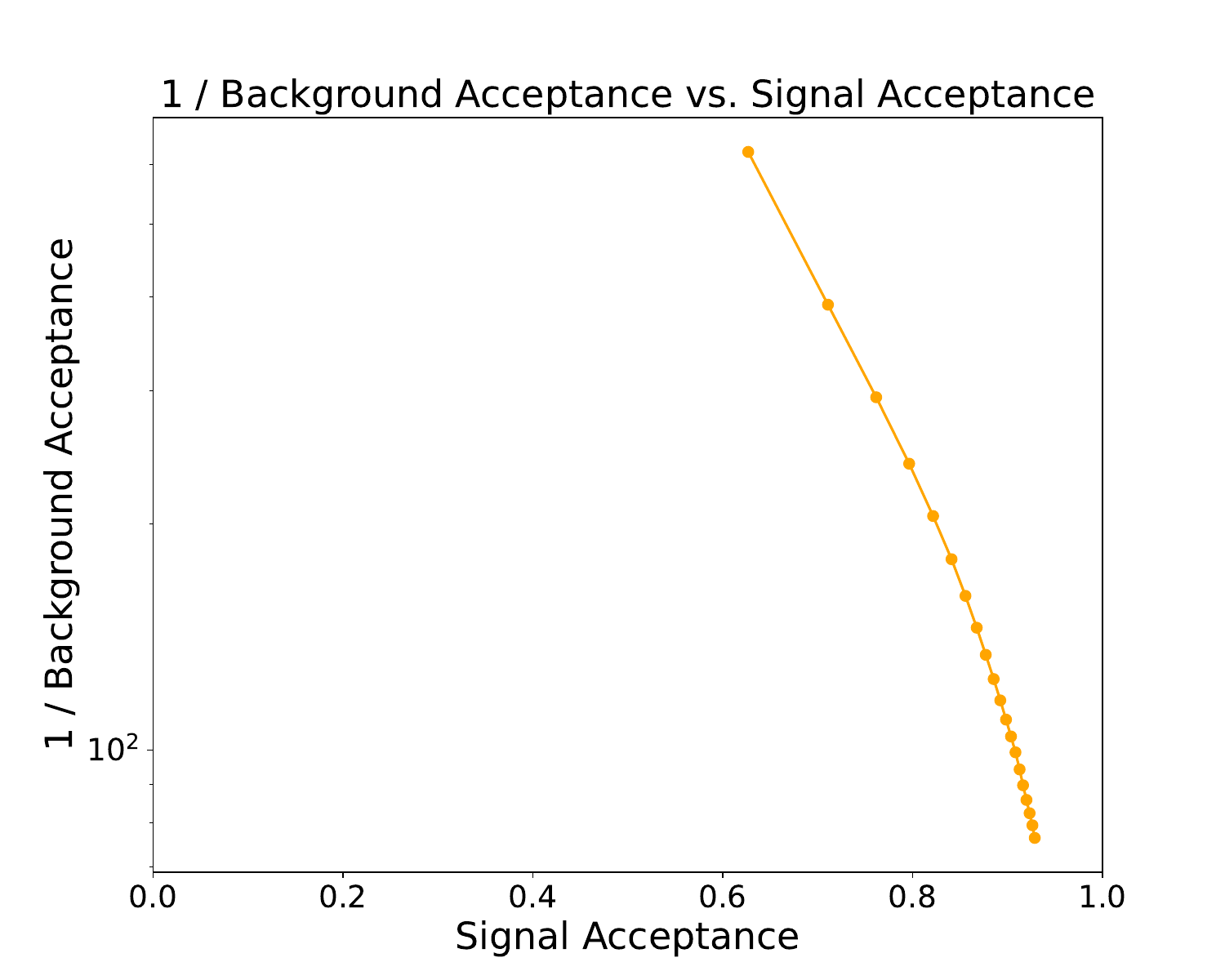}}
	\caption{1$/$Background Acceptance against the Signal Acceptance for $h \to b\bar{b}$ channel. (a) $M_{H_F} = 800$ GeV, (b) $M_{H_F} = 900$ GeV, (c) $M_{H_F} = 1000$ GeV, (d) $M_{H_F} = 1100$ GeV, (e) $M_{H_F} = 1200$ GeV, (f) $M_{H_F} = 1300$ GeV and (g) $M_{H_F} = 1400$ GeV.}
	\label{fig:Roc_hbb}
\end{figure}

%The BDT training is performed using the MC-simulated samples. 
The BDT training is performed using the MC-simulated samples. The number of MC events are a dataset consisting of 200000 signal events and the samples of background events described above, each containing 200000 events. The data is split as follows: training dataset is 70\% of the total events, while the testing dataset is 30\% of the total events. To optimize the BDT model, we use \textit{HyperOpt}\cite{Hiper:2013} to tune the hyperparameters of the classifier.
As an example of the performances of the trainings, Fig.~\ref{fig:Roc_hbb} shows the plots for 1$/$Background Acceptance against the Signal Acceptance for the $h \to b\bar{b}$ channel.
%As an example of the performances of the trainings, we present in Fig.~\ref{fig:Roc_hbb} the Receiver Operating Characteristic (ROC) curve for the $h \to b\bar{b}$ channel.

These signal and background samples are scaled to the expected number of candidates, which is calculated based on the integrated luminosity and cross-sections. The BDT selection is optimized individually for each channel to maximize the figure of merit, i.e., the \textit{signal significance}, defined as $S/\sqrt{S+B}$, where $S$ and $B$ represent the number of signal and background candidates, respectively, in the signal region after applying the selection criteria. 
		
\begin{figure}[!htb]
			\begin{center}
				\subfigure[]{\includegraphics[scale=0.35]{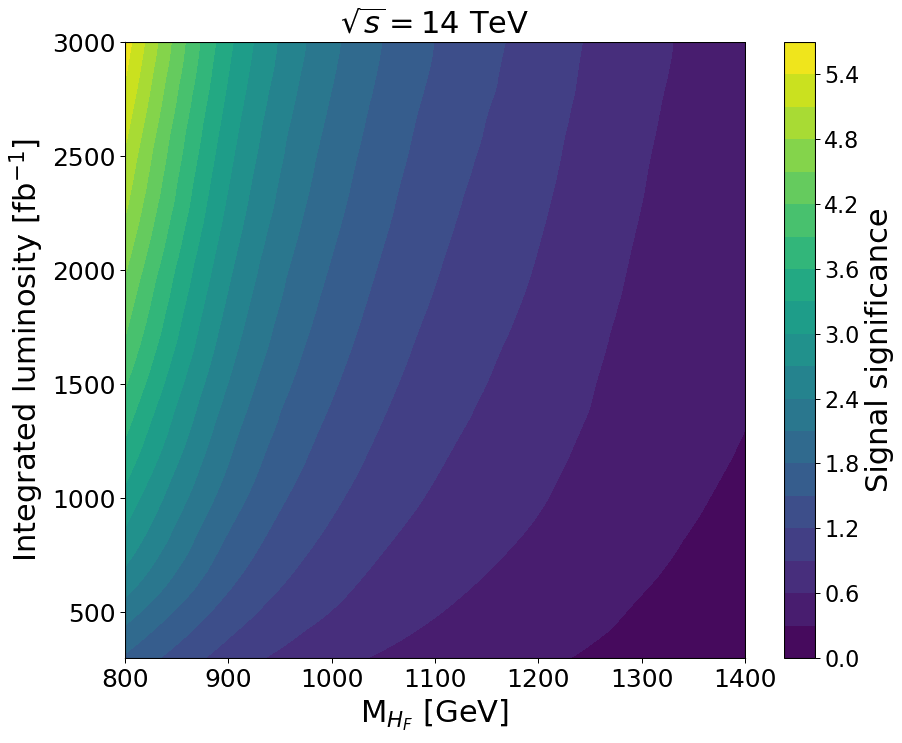}}
				\subfigure[]{\includegraphics[scale=0.35]{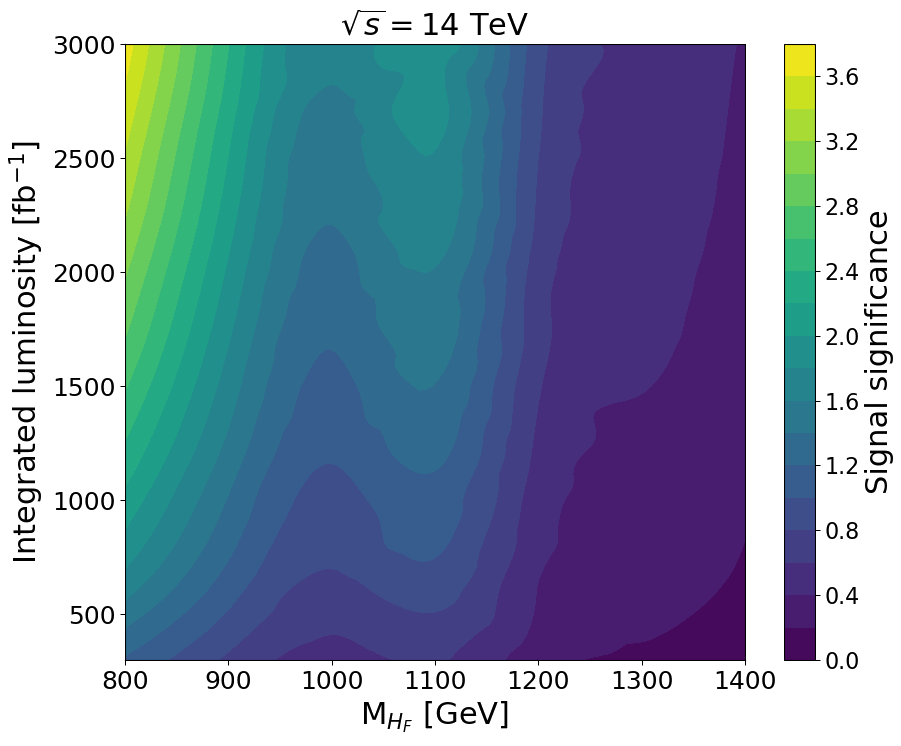}}
				\subfigure[]{\includegraphics[scale=0.35]{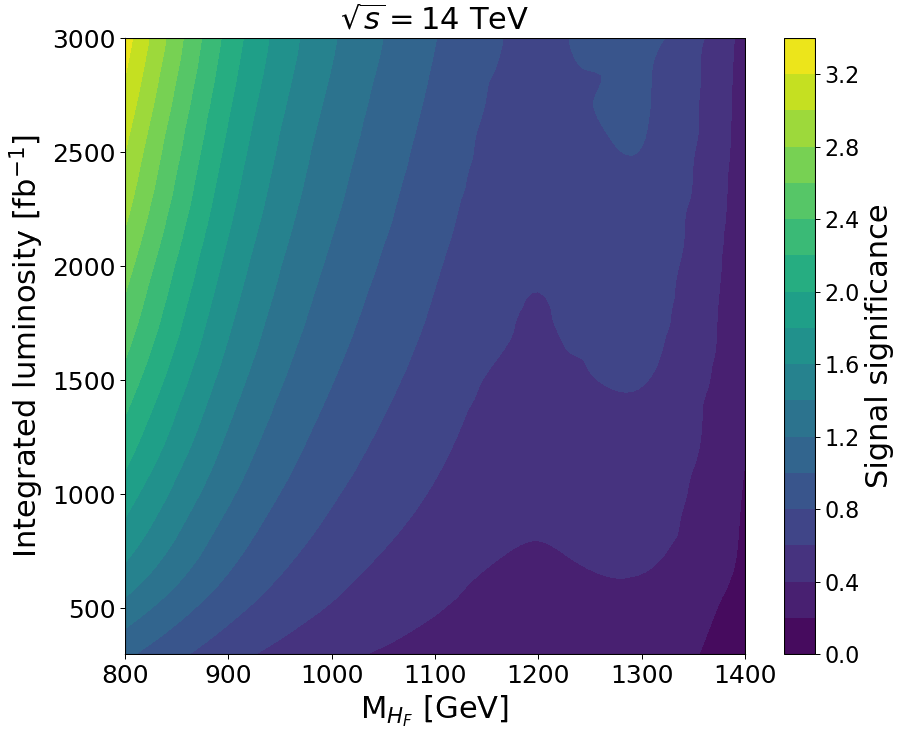}}
			\end{center}
			\caption{Density plot showing the \textit{signal significance} for the $h\to b\bar{b}$ channel as a function of the integrated luminosity and the flavon mass $M_{H_F}$: (a) $\textbf{S1}$, (b) $\textbf{S2}$ and (c) $\textbf{S3}$.}
			\label{fig:Hbbh_SigplotS1}
		\end{figure}

		In Fig. \ref{fig:Hbbh_SigplotS1} we show the \textit{signal significance} (for the three scenarios $\textbf{Si}$) as a function of the luminosity and the flavon mass $M_{H_F}$ for the $pp \to H_F\to h b \bar{b}(h\to b \bar{b})$ process.
		We observe that at a 14 TeV $pp$ collider with an integrated luminosity of 3000 fb$^{-1}$, strong sensitivity exists for the various choices of $M_{H_F}$. We found that for the range $800<M_{H_F}<950$ GeV a \textit{signal significance} between $3-5.6\sigma$ was obtained. Thus, with the arrival of the future HL-LHC, the flavon particle could be discovered if the scenario $S1$ is the one chosen by nature. For scenarios $\textbf{S2}$ and $\textbf{S3}$ there is a depletion around 1.2 TeV, and then the significance increases around 1.3 TeV, this can be explained by the slightly better performance of the BDT for the 1.3 TeV mass regarding 1.2 TeV. 
		
\begin{figure}[!htb]
			\begin{center}
				\subfigure[]{\includegraphics[scale=0.35]{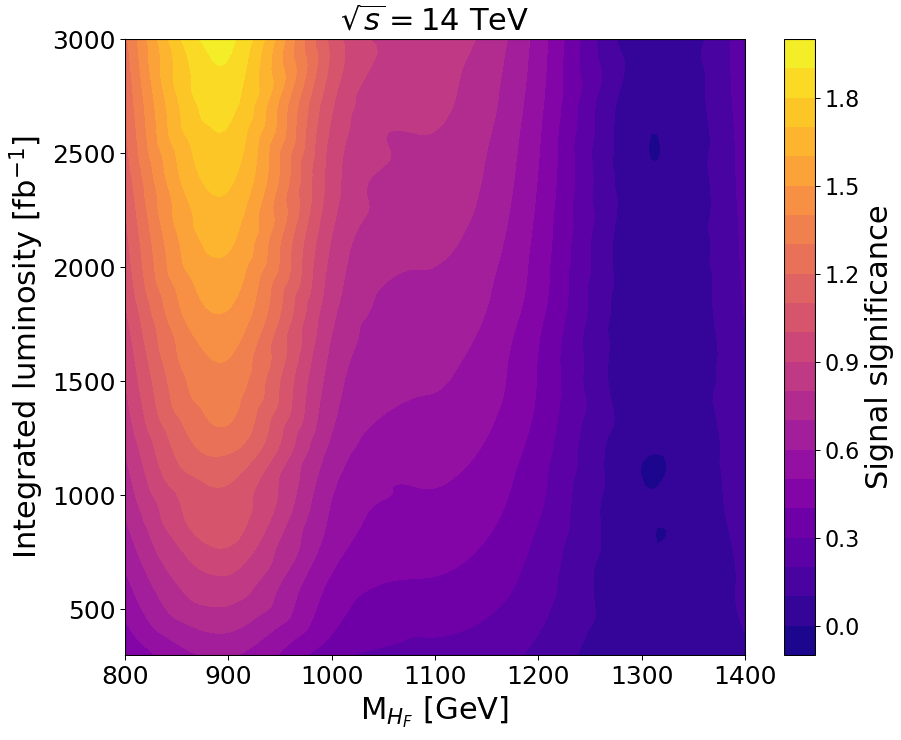}}
				\subfigure[]{\includegraphics[scale=0.35]{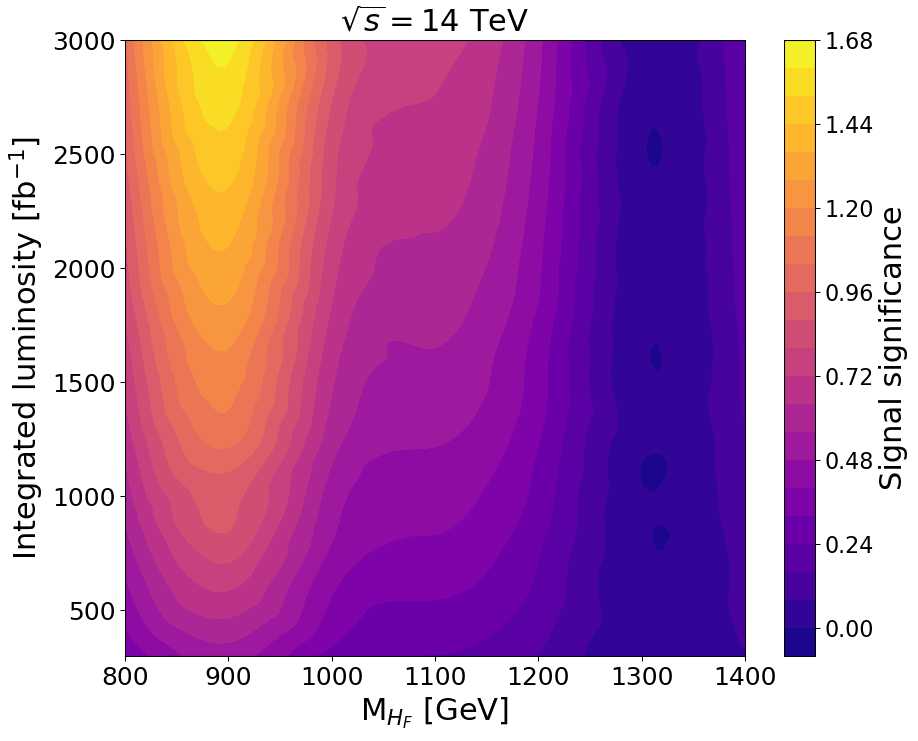}}
				\subfigure[]{\includegraphics[scale=0.35]{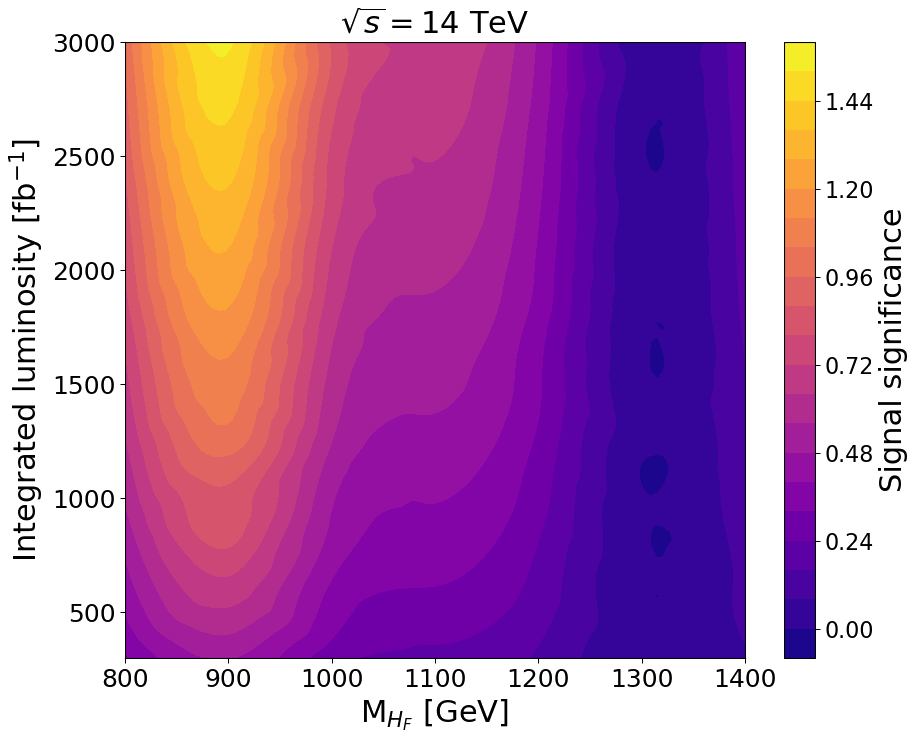}}
			\end{center}
			\caption{Density plot showing the \textit{signal significance} for the $h\to \gamma\gamma$ channel as a function of the integrated luminosity and the flavon mass $M_{H_F}$: (a) $\textbf{S1}$, (b) $\textbf{S2}$ and (c) $\textbf{S3}$.}
			\label{fig:Haabb_SigplotS1}
		\end{figure}

		Meanwhile, Fig. \ref{fig:Haabb_SigplotS1} shows the \textit{signal significance} as a function of luminosity and the flavon mass $M_{H_F}$ for the $pp \to H_F\to h b \bar{b}(h\to \gamma \gamma)$ process. In this channel, the sensitivity is lower than in the previous one because there is a factor of $10^{-3}$ that suppresses the previous case, such a factor comes from the $\mathcal{BR}(h\to \gamma\gamma)\sim 10^{-3}$. However, considering that a future 100 TeV collider is a possibility \cite{HL-lhc2}, it is important to note that our calculations estimate that for a luminosity of 30 ab$^{-1}$, the sensitivity could reach up to 5$\sigma$.
		
	Finally, to highlight the usefulness of BDT, in Table $\text{\ref{tab:Signal-significance_StragCUTS}}$
	we present the \textit{signal significance} for the $h\to b\bar{b}$ channel,
	where we have considered the basic cut, included in the HL-LHC card~\cite{HL-LHC_card},
	and the following kinematic cuts for $M_{H_F}=800$ GeV:
	\begin{enumerate}
		\item $p_{T}(b_{1,\,2,\,3,\,4})>80,\,60,\,50,\,25$ GeV,
		\item $650<M_{{\rm inv}}(b_{1}b_{2}b_{3}b_{4})<900$ GeV,
		\item Combined transverse momentum $p_{T}(b_{1}b_{2}b_{3}b_{4})>550$ GeV.
	\end{enumerate}
	\begin{table}
		
		\caption{Signal significance for the channel $h\to b\bar{b}$ and $\mathcal{{L}_{{\rm int}}}$=3000
			fb$^{-1}.$\label{tab:Signal-significance_StragCUTS} }
		
		\begin{centering}
			\begin{tabular}{|c|c|}
				\hline 
				$M_{H_{F}}$(GeV) & Signal Significance\tabularnewline
				\hline 
				\hline 
				800 & 1.75 $\sigma$\tabularnewline
				\hline 
				900 & 0.87$\sigma$\tabularnewline
				\hline 
				1000 & 0.4872$\sigma$\tabularnewline
				\hline 
			\end{tabular}
			\par\end{centering}
	\end{table}
Therefore, we conclude that the use of BDT substantially improves the isolation of the proposed signals.
		
		%%%%%%%%%%%%%%%%%%%%%%%%%%%%%%%%%%%%%%%%%%%%%
		%%%%%%%%%%%%%%%%%%%%%%%%%%%%%%%%%%%%%%%%%%%%%
		%%%%%%%%%%%%%%%%%%%%%%%%%%%%%%%%%%
		\section{Conclusions}\label{se:concl}
		In this paper we studied the possibility of observing the $pp\to H_F\to h b\bar{b}$ process through two channels in which the Higgs boson decays, namely, $h\to b\bar{b}$ and $h\to \gamma\gamma$. The $H_F$ mother particle is a predicted state in a model that extends the Standard Model and incorporates the Froggatt-Nielsen mechanism which is an opportunity to explain the mass hierarchy. Our predictions consider realistic scenarios for the free parameters of the model, which were extracted by analyzing theoretical and experimental constraints. We identify specific regions of the model parameter space in which we found \textit{signal significances} of up to $5\sigma$ ($2\sigma$) for the $h\to b\bar{b}$ ($h\to \gamma\gamma$) channel. The study was based on the High Luminosity Large Hadron Collider which aims $3000$ fb$^{-1}$ of accumulated data at $14$ TeV. 
		%%%%%%%%%%%%%%%%%%%%%%%%%%%%%%%%%%
		%%%%%%%%%%%%%%%%%%%%%%%%%%%%%%%%%
		%%%%%%%%%%%%%%%%%%%%%%%%%%%%%%%%%%
		%%%%%%%%%%%%%%%%%%%%%%%%%%%%%%%%%
		%%%%%%%%%%%%%%%%%%%%%%%%%%%%%%%%%%
		%%%%%%%%%%%%%%%%%%%%%%%%%%%%%%%%%

		\subsection*{Acknowledgments}
		The work of Marco A. Arroyo-Ure\~na and T. Valencia-P\'erez is supported by ``Estancias posdoctorales por M\'exico (SECIHTI)" and ``Sistema Nacional de Investigadores" (SNII-SECIHTI). T.V.P. acknowledges support from the UNAM project PAPIIT IN111224 and the CONAHCYT project CBF2023-2024-548. E. A. Herrera-Chac\'on acknowledges support from CONAHCYT.

		\bibliographystyle{utphys}
		\bibliography{REFc}

	\end{document}